\documentclass[]{aa}   
\usepackage{graphicx}
\usepackage{multirow}
\usepackage{subcaption}
\usepackage[normalem]{ulem}

\usepackage{color}

\definecolor{tom}{rgb}{0.0, 0., 1.0}

\definecolor{lianne}{rgb}{0.7, 0.0, 0.5}

\makeatletter
\newcommand*{\rom}[1]{\expandafter\@slowromancap\romannumeral #1@}
\makeatother

\usepackage{siunitx}
\newcommand{\Alfvenic}{Alfv\'enic }
\newcommand{\Alfven}{Alfv\'en }
\usepackage[utf8]{inputenc}
\usepackage{fourier} 
\usepackage{array}
\usepackage{makecell}

\usepackage{placeins}

\begin{document}
\title{Forward Modelling of MHD Waves in Braided Magnetic Fields}
\author{L.~E. Fyfe \inst{1} \and T.~A. Howson \inst{1} \and I. De Moortel \inst{1, 2}}
\institute{School of Mathematics and Statistics, University of St. Andrews, St. Andrews, Fife, KY16 9SS, U.K. \and Rosseland Centre for Solar Physics, University of Oslo, PO Box 1029  Blindern, NO-0315 Oslo, Norway}

\abstract{}
{We investigate synthetic observational signatures generated from numerical models of transverse waves propagating in braided magnetic fields.}
{We examine two simulations with different levels of magnetic field braiding and impose a periodic, transverse wave driver at the lower boundary. After waves reflect off the top boundary, a complex pattern of wave interference forms. We analyse the synthetic emission produced by the forward modelling code, FoMo. We examine line intensity, Doppler shifts and kinetic energy along several line-of-sight (LOS) angles.}
{The Doppler shifts showed the presence of transverse waves. However, when analysing the intensity, waves are less easily observed for more complex magnetic fields and may be mistaken for background noise. Depending on the LOS angle, the observable signatures of waves reflect some of the magnetic field braiding, particularly when multiple emission lines are available. In the more braided simulation, signatures of phase mixing can be identified. We identify possible ambiguities in the interpretation of wave modes based on the synthetic emission signatures.}
{Most of the observables within this article behave as expected, given knowledge of the evolution of the parameters in 3D space. However, some intriguing observational signatures are present. Detecting regions of magnetic field complexity is somewhat possible when waves are present. However, simultaneous spectroscopic imaging from different lines is important to identify these locations. Care needs to be taken when interpreting intensity and Doppler velocity signatures as torsional motions as, in our setup, such signatures were a consequence of the magnetic field complexity and not torsional waves. Finally, the kinetic energy, estimated with Doppler velocities, is dependent on the polarisation of the wave, complexity of the background field and the LOS.}

\keywords{Sun: corona - Sun: magnetic fields - Sun: oscillations - magnetohydrodynamics (MHD)}
\maketitle


\section{Introduction}\label{sec:introduction}

Over the past couple of decades, the existence of MHD waves throughout the solar atmosphere has become more apparent, due to higher spatial and temporal resolution in imaging and spectroscopic instruments \citep[e.g.][]{DeMoortelNakariakov2012,  Arregui2015, VanDoorsselaereNakariakov2020}. Within the solar corona, an assortment of these waves have been identified. Particularly pertinent to this article is the presence of transverse waves propagating along coronal magnetic structures \citep[e.g.][]{AschwandenFletcher1999, AschwandenDePontieu2002, OkamotoTsuneta2007,TomczkMcIntosh2007,McIntoshDePontieu2011,ThurgoodMorton2014,AnfinogentovNakariakov2015, MortonTomczyk2015, TianYoung2016, DuckenfieldAnfinogentov2018}.

For quite some time, MHD waves have been proposed as a mechanism to transfer energy into the solar corona, and by dissipation of this energy, help maintain the high coronal temperatures. Further details of coronal heating theories can be found in the reviews by, for example, \citet{WalshIreland2003, Klimchuk2006, ParnellDeMoortel2012, DeMoortelBrowning2015, Klimchuk2015}. 

MHD waves can be generated by short time scale motions located at photospheric magnetic footpoints \citep[e.g.][]{CranmerBallegooijen2005,WangOfman2009, Hillier2013}, however motions with longer time scales result in braiding, and hence stressing, of the coronal magnetic field \citep[e.g.][]{Parker1972}. As a result, it is thought that the coronal magnetic field is constantly in a complex configuration. The combination of footpoint motions and intricate topological magnetic fields will enhance the formation of current sheets, leading to heating \citep[e.g.][]{LongcopeSudan1994, LongbottomRickard1998, PeterGudiksen2004, WilmotSmmithPontin2011, WilmotSmith2015, OHaraDeMoortel2016, PontinCandelaresi2016, RealeGuarrasi2016}. In addition, numerical simulations have shown that in loops that have been twisted sufficiently by such (slow) footpoint motions, the onset of the kink instability can release the stored magnetic energy, leading to heating events \citep[e.g.][]{BrowningGerrard2008, HoodBrowning2009, ReidHood2018}. Further studies have gone on to forward model such kink instabilities and determine whether there are any observational signatures from such phenomenon \citep[e.g.][]{SnowBotha2017}. 

Phase mixing has been proposed as a mechanism to increase the rate of heating due to (Alfv\'en) wave dissipation. This is achieved by transverse gradients in the Alfv\'en speed profile causing large spatial gradients in the velocity and magnetic fields \citep{HeyvaertsPriest1983}. However, concerns have been highlighted \citep[e.g][]{CargillDeMoortel2016} as to whether this model can self-consistently deliver the heating required to balance thermal losses on the right timescales. In particular, \cite{CargillDeMoortel2016} show that much larger transport coefficients than expected in the corona would be required \citep[see also e.g.][]{PaganoDeMoortel2017, PaganoPascoe2018, PaganoDeMoortel2019}. 

Waves are also important in the field of coronal seismology \citep[e.g][]{Uchida1970, RobertsEdwin1984, TaroyanErdelyi2009, ChenPeter2015, PascoeAnfinogentov2018, KaramimehrVasheghaniFarahani2019, PascoeHood2019, WangOfman2019, MagyarNakariakov2020} i.e. estimating plasma parameters using the characteristics of observed waves and oscillations. Hence, there is a clear need for a better understanding of MHD wave dynamics in realistic coronal fields. For this reason, previous authors have investigated the signatures of such waves (for example sausage, slow and the kink modes) with the use of synthetic observables generated from numerical models \citep[e.g.][]{AntolinVanDoorsselaere2013, YuanVanDoorsselaere2015, YuanVanDoorsselaere2016}.

In this study, we investigate the observables from synthetic emission data produced by two numerical simulations of transverse MHD waves in coronal plasma. Each experiment considers different degrees of magnetic field complexity \citep{HowsonDeMoortel2020}. We examine the effect the magnetic field has on the observables and determine whether there are any distinguishable signatures. This will help towards cataloguing potential observables for detecting similar MHD waves and will help close the gap between numerical simulations and observations. Unlike many previous articles, which study waves in simple cylindrical models, this paper investigates waves in a more complex coronal environment. In Sect. \ref{sect_numerical_model}, we give a brief overview of the numerical model and results of \citet{HowsonDeMoortel2020}. In Sect. \ref{forward_modelling_section}, we analyse the synthetic emission data by examining the imaging and spectral signatures. Our findings are then discussed and summarised in Sect. \ref{sec_Discussion}.

\section{Numerical model} \label{sect_numerical_model}
\subsection{Setup}

We begin with a brief description of the numerical model for which we generate the synthetic emission data discussed in this article (see Sect. \ref{forward_modelling_section}). For further details and analysis of the simulation results, we refer the reader to \citet{HowsonDeMoortel2020}. 

\begin{figure}[t!]
  \centering
  \includegraphics[width=0.5\textwidth]{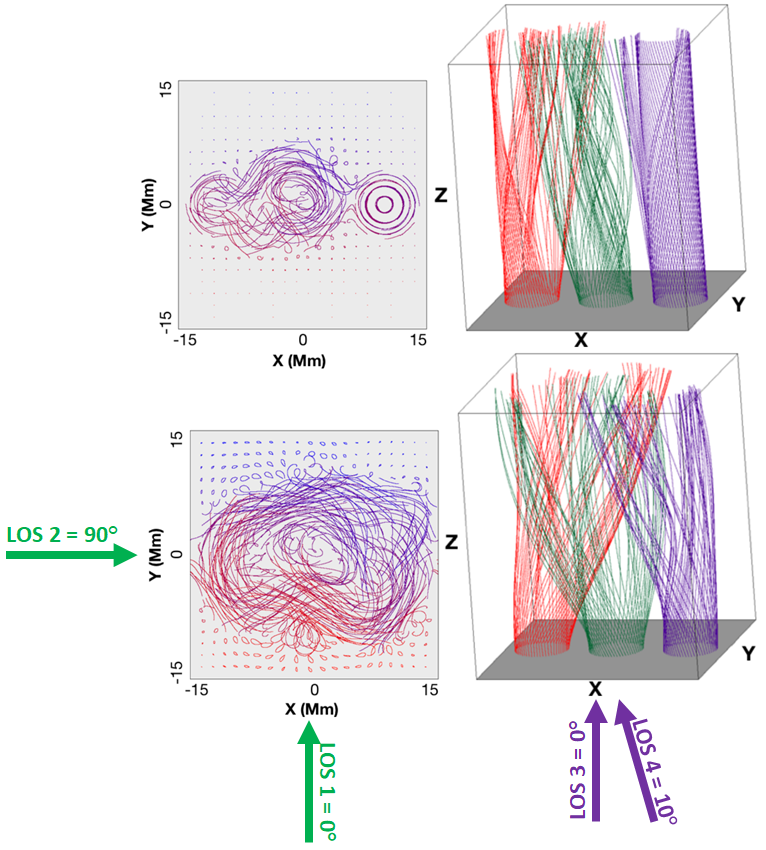}
  \caption{Illustration of the magnetic field line complexity for IC1 (upper row) and IC2 (lower row) showing the projection of the magnetic field lines onto the \(xy\)-plane (left column) and the configuration of representative field lines in 3D (right column), modified from \citet{HowsonDeMoortel2020}. The LOS angles (LOS1, 2, 3 and 4), used in the forward modelling, are green arrows if they are in the \(xy\)-plane (left) and purple in the \(xz\)-plane (right).}
  \label{los_angles}
\end{figure}

To investigate the effects of complex magnetic field structures on \Alfvenic waves, \citet{HowsonDeMoortel2020} consider two initial conditions, each derived from different times within simulations performed by \citet{ReidHood2018}. In the latter study, the authors examine the behaviour of an avalanche model in three twisted magnetic threads. These threads are twisted by counter rotational boundary drivers located at each of their footpoints. The three threads are rotated at different rates, such that the central thread becomes kink unstable first. This subsequently destabilises each of the neighbouring threads. Intricate current structures are generated during this process, predominantly in \(j_z\) (the initial field is aligned with the $z$-axis). The two snapshots used in \citet{HowsonDeMoortel2020} are a time when one of the threads still has a recognisable structure and later on, when any coherent structuring has been lost. We will denote these initial conditions as IC1 and IC2, respectively (Fig \ref{los_angles}). A more in-depth analysis of the avalanche model can be found in \cite{ReidHood2018}. In \citet{HowsonDeMoortel2020}, these initial conditions are then relaxed numerically by switching off the footpoint driving and implementing a large viscosity to reduce the amplitude of any flows and oscillations. The numerical relaxation is stopped once the magnitude of any remaining velocities is small compared to the amplitude of the wave driving (see below).

\begin{table}[t!]
    \centering
    \makebox{\begin{tabular}{|| m{1cm} | m{3cm} | m{2cm} ||}
        \hline\hline
         \thead{LOS} & \thead{Parallel To} & \thead{POS}  \\
        \hline\hline
         \makecell{1}  & \makecell{positive \(y\)-axis} & \makecell{\(xz\)-plane} \\
        \hline
          \makecell{2}  & \makecell{positive \(x\)-axis} & \makecell{\(yz\)-plane} \\
         \hline
          \makecell{3}  & \makecell{positive \(z\)-axis} & \makecell{\(xy\)-plane} \\
         \hline
          \makecell{4}  & \multicolumn{2}{c||}{\makecell{\ang{10} off LOS3 as illustrated in Fig. \ref{los_angles}.}} \\
         \hline\hline
    \end{tabular}}
    \caption{LOS angles given in reference to their plane of the sky (POS) and the axis they are parallel to. These are also depicted in Fig. \ref{los_angles}.}
    \label{los_angle_table}
\end{table}

The numerical simulations in \citet{HowsonDeMoortel2020} and \citet{ReidHood2018} use the Lagrangian-remap code, Lare3D \citep{ArberLongbottom2001}, which solves the fully 3D normalised non-ideal MHD Equations. In \citet{HowsonDeMoortel2020}, the effects of thermal conduction, optically thin radiation and gravity were neglected. The background viscosity was set to zero, however, to ensure numerical stability, the Lare3D code does include shock viscosity. This contributes to both the viscous force and viscous heating term. The simulations were non-resistive (\({\eta = 0}\)) and hence, the only increases in temperature are due to compression and the (small) shock viscosity heating term.

We will refer to the numerical simulations of \citet{HowsonDeMoortel2020} as S1 and S2 (corresponding to initial conditions IC1 and IC2, respectively). The numerical domain has dimensions of \(\text{30 Mm}\times \text{30 Mm} \times \text{100 Mm}\), using a numerical grid of \({256 \times 256 \times 1024}\) cells.

The transverse waves investigated in \citet{HowsonDeMoortel2020} are excited by imposing a wave driver on the lower \(z\) boundary of the form \(\textbf{v}\left(t\right)=\left(0,v_y,0\right)\), where \(v_y\) is defined by

\begin{equation*}
 \ v_y(t) = v_0\text{ sin}\left(\omega t\right),
 \label{wave_driver_equn} 
 \end{equation*}
 
 \noindent with an amplitude, \(v_0 \approx 20 \text{ km }\text{s}^{-1}\) and frequency, \(\omega \approx 0.21\text{ s}^{-1}\). This corresponds to a wave period of \(\tau \approx 28\text{ s}\).  We have implemented a relatively high frequency driver to allow multiple wavelengths to fit within the length of the numerical domain (100 Mm). Although the majority of wave or oscillatory periods observed in the solar corona are of the order of a few minutes \citep[e.g.][]{TomczkMcIntosh2007, TomczykMcIntosh2009, MortonTomczyk2016}, observations of wave periods of a few tens of seconds have also been identified along coronal loops \citep[e.g.][]{WilliamsPhillips2001} and spicules \citep[e.g.][]{ OkamotoDePontieu2011, YoshidaSuematsu2019}.
 
 The \(x\) and \(y\) boundaries are periodic and the \(z\) boundaries set the gradients of all the variables to be zero, with the exception of the velocity, where the wave driver is imposed on the bottom boundary and the velocity is set to zero on the top. This ensures that waves are reflected at the top boundary and forced to propagate back into the domain.
 


\subsection{Evolution} \label{evolution}

In order to interpret the synthetic emission data generated by the forward modelling (see Sect. \ref{forward_modelling_section}), it is helpful to first briefly discuss the evolution of the system, in particular the behaviour of the velocity, temperature and density.

 \begin{figure}[bp!]
\centering
\vspace{0cm}
\begin{subfigure}{0.25\textwidth}
  \centering
  \hspace{0cm}
  \makebox[0pt]{\includegraphics[width=1.\textwidth]{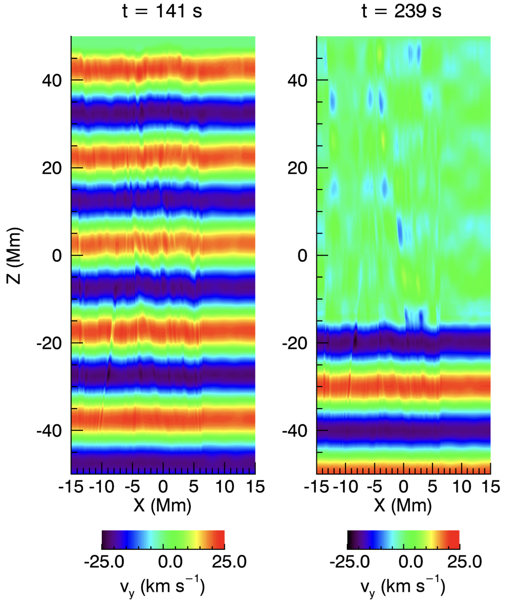}}
  \caption{}
  \label{tom_vy_s1}
\end{subfigure}%
\begin{subfigure}{0.25\textwidth}
  \centering
  \makebox[0pt]{\includegraphics[width=1\textwidth]{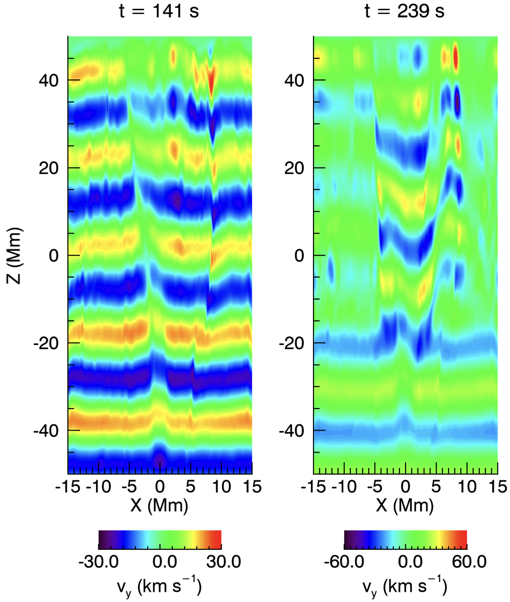}}
  \caption{}
  \label{tom_vy_s2}
\end{subfigure}
\caption{Contour plot of \(v_y\) in a vertical slices through \(y=0\) Mm at 141 s (first panels) \& 239 s (second panels) for simulations (a) S1 and (b) S2. Note the change in the range of the colour bars.}
\end{figure}

\begin{figure}[b!]
\centering
\includegraphics[width=0.4\textwidth]{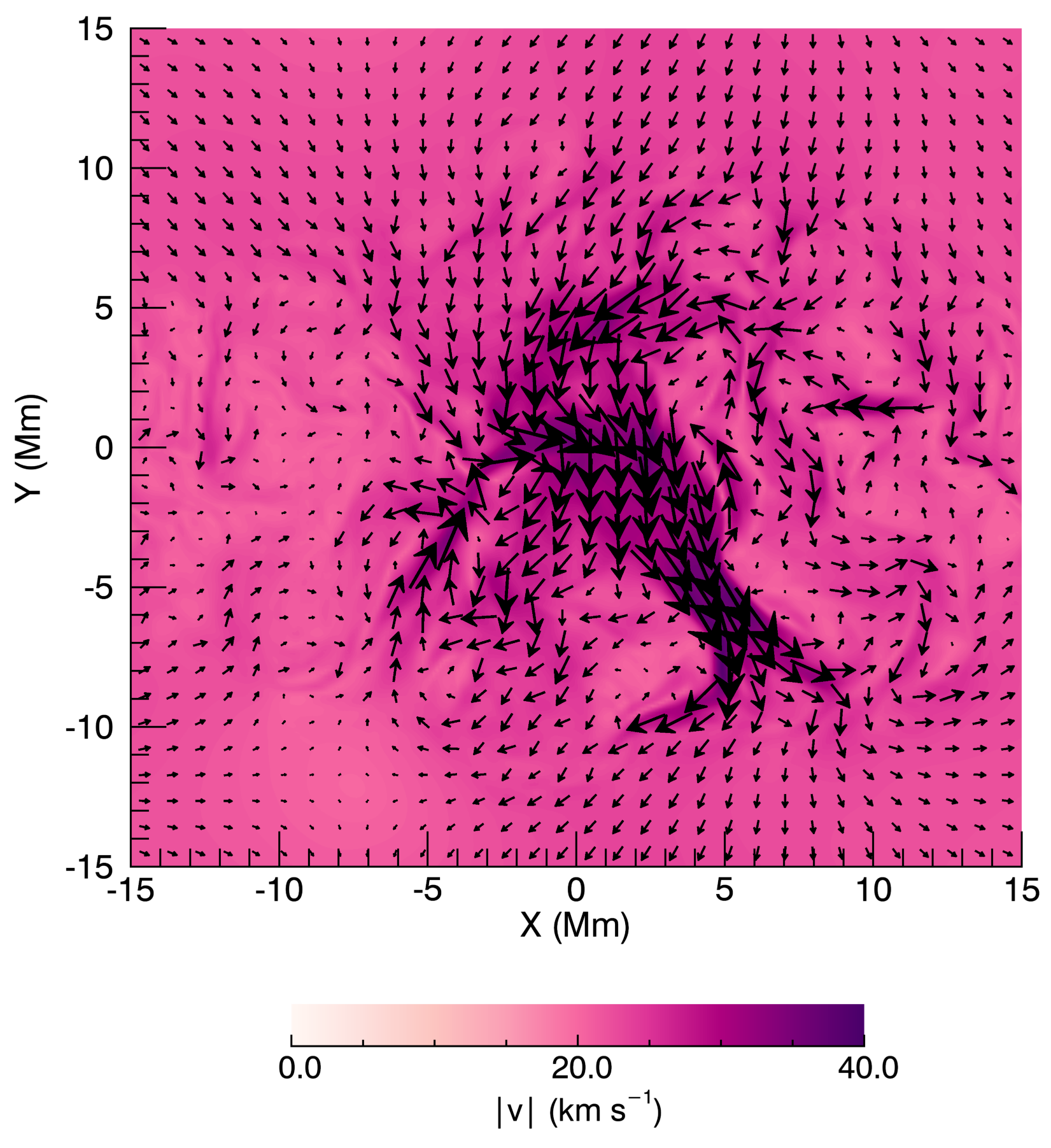}
\caption{Illustration of the speed (contours) and the horizontal velocity (vectors) through the horizontal midplane for S2 at \(\text{t } = 239 \text{ s}\).}
\label{tom_vh_s2}
\end{figure}

One significant feature seen in \citet{HowsonDeMoortel2020} was the presence of phase mixing, where the formation of small length scales depends on the amount of magnetic field complexity. The left-hand panels of Figs. \ref{tom_vy_s1} \& \ref{tom_vy_s2} are snapshots of $v_y$ before the first reflection off the top boundary. They each show the distortion of the (velocity) wave front as it propagates through the complex magnetic field. The degree of distortion clearly relates to the amount of braiding (complexity) of the magnetic field, i.e.~the more complex the field the greater the level of phase mixing. The last two panels in Figs. \ref{tom_vy_s1} and \ref{tom_vy_s2} illustrate the behaviour of the waves at a later time (t = 239s), when they have reflected back into the domain and begin to generate wave interference.

Although the imposed boundary driver is incompressible, non-linearity and coupling to fast modes lead to compressibility as the waves propagate through the domain. Figure \ref{tom_vh_s2} illustrates the complex and compressible nature of the velocity field through a horizontal slice at the midplane in S2. Clearly, the velocity is no longer aligned with the boundary wave driver. This (partial) change to the polarisation of the wave, from \(v_y\) to \(v_x\), is due to the interaction of the perturbations in \(B_y\) with \(j_z\), which produces an \(x\)-component of the  Lorentz force \citep[see also][]{HowsonDeMoortel2019} and hence, generates velocity perturbations in the \(x\)-direction, as the waves travel along the twisted magnetic field lines.

\begin{figure*}
\centering
\vspace{0cm}
\begin{subfigure}{1.\textwidth}
  \centering
  \hspace{0cm}
  \includegraphics[width=0.8\textwidth]{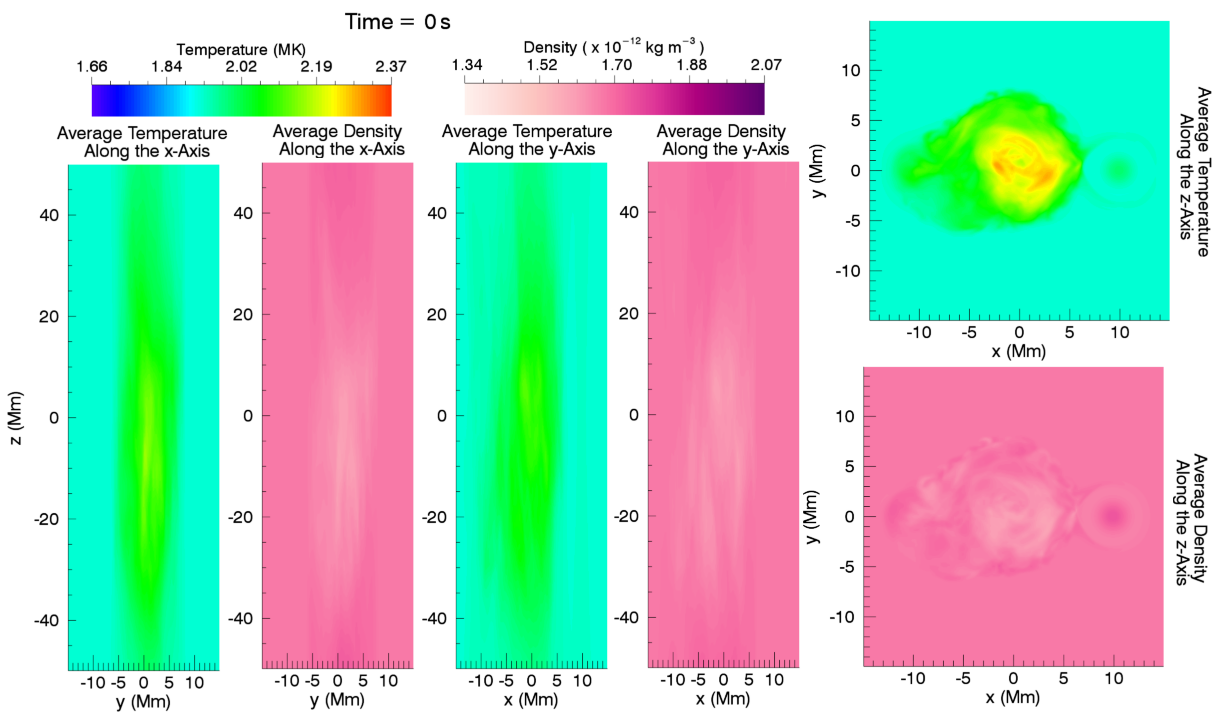}
  \caption{}
  \label{v_temp_rho_time20_0}
\end{subfigure}
\begin{subfigure}{1.\textwidth}
  \centering
  \hspace{0cm}
\includegraphics[width=0.8\textwidth]{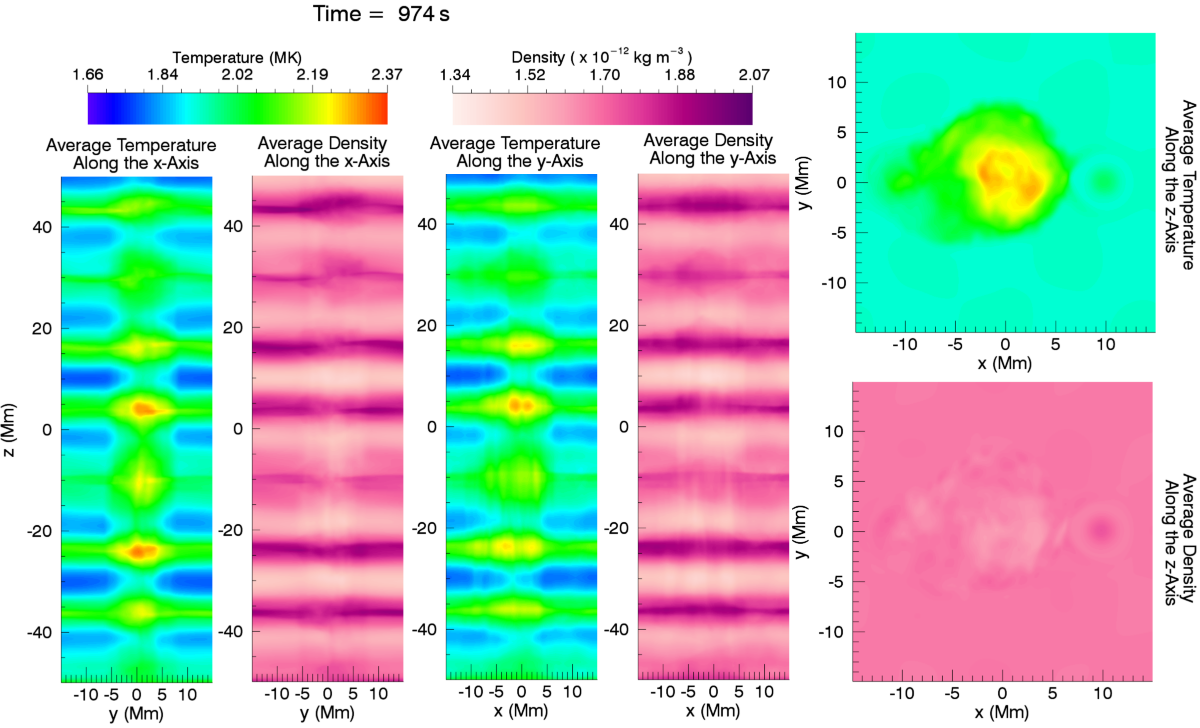}
  \caption{}
  \label{av_temp_rho_time20_974}
\end{subfigure}
\caption{Contours of average temperature (rainbow) and density (pink) along the three axes at (a) 0s and (b) 974s for S1.}
\label{temp_rho_contours_time20}
\end{figure*}

\begin{figure*}
\centering
\vspace{0cm}
\begin{subfigure}{1.\textwidth}
  \centering
  \hspace{0cm}
  \includegraphics[width=0.8\textwidth]{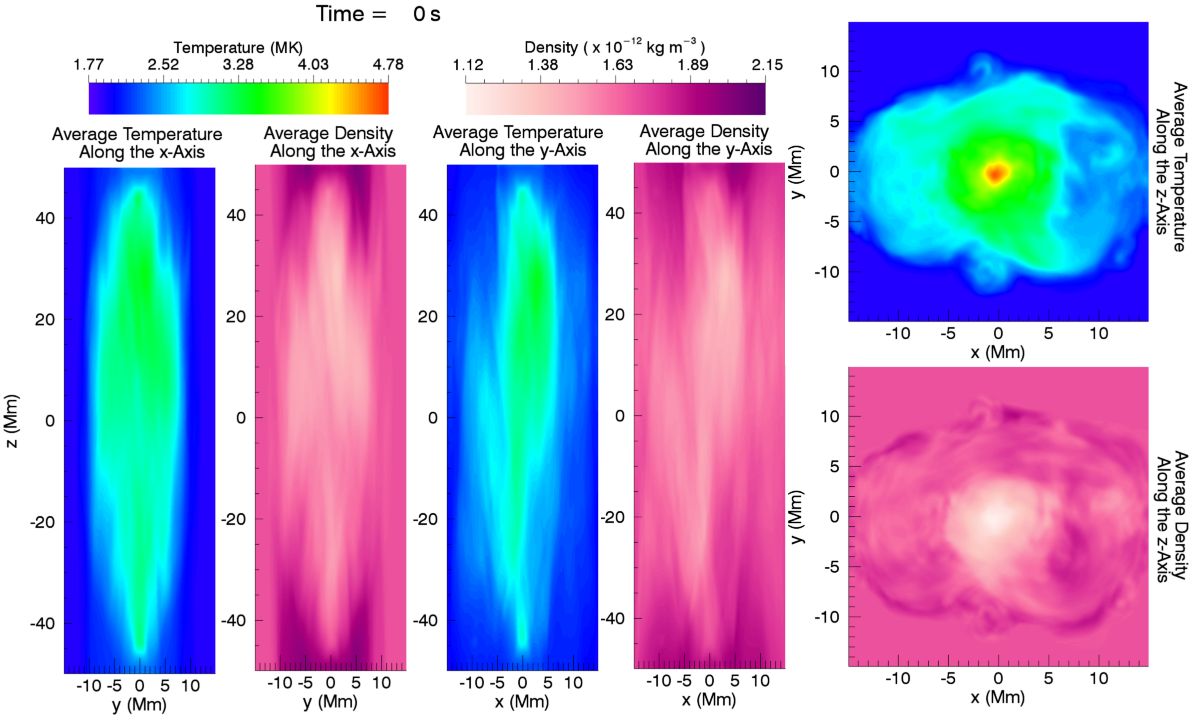}
  \caption{}
  \label{v_temp_rho_time100_0}
\end{subfigure}
\begin{subfigure}{1.\textwidth}
  \centering
  \hspace{0cm}
\includegraphics[width=0.8\textwidth]{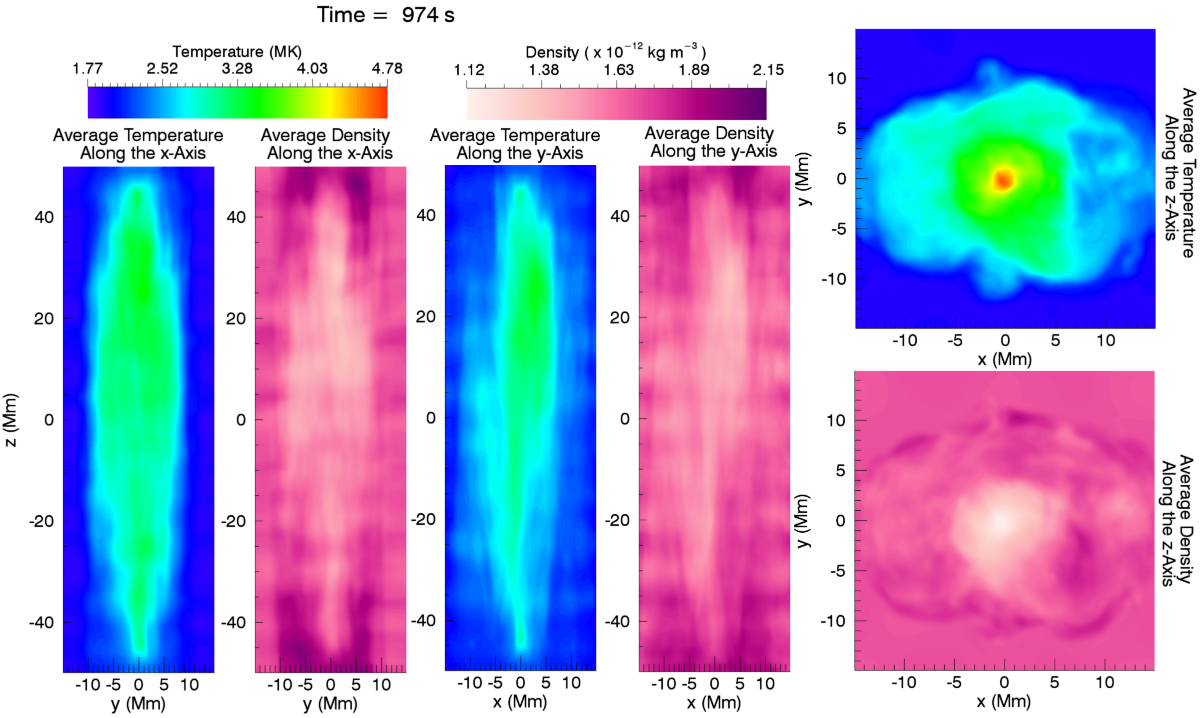}
  \caption{}
  \label{av_temp_rho_time100_974}
\end{subfigure}
\caption{Contours of average temperature (rainbow) and density (pink) along the three axes at (a) 0s and (b) 974s for S2.}
\label{temp_rho_contours_time100}
\end{figure*}

The density and temperature profiles averaged along the three main axes are illustrated in Figs. \ref{temp_rho_contours_time20} and \ref{temp_rho_contours_time100} for simulations S1 and S2, respectively (where the scale of the colour bar changes between S1 and S2). Initially, a tube-like structure is present in both simulations when viewing along the \(x\) and \(y\)-axes. These are regions of higher temperature, compared to the surroundings, and occur where the viscous and Ohmic heating is largest in \citet{ReidHood2018}. Due to pressure balance and the larger temperatures, there is a drop in density within the tube-like structures \citep[see][]{HowsonDeMoortel2019}. 
Therefore, in this case, the \Alfven speed is generally higher inside the magnetic structures than outside. This is converse to what is typically used in wave studies and is more common for structures in the lower atmosphere but has also been studied in a coronal context \citep[e.g.][]{HowsonDeMoortelAntolin2019}. Although in S1 the right-hand thread is a distinct structure (see Fig.\ref{los_angles}), this is only apparent in the temperature and density averaged along the $z$-axis and is not evident when averaging along the \(x\) \& \(y\)-axes.

As the simulations evolve, the compressible nature of the velocity field becomes evident in Figs. \ref{av_temp_rho_time20_974} and \ref{av_temp_rho_time100_974}, where we see perturbations in both the density and temperature along the \(x\) \& \(y\)-axes. As discussed above, in the initial equilibria, regions of high temperature typically coincide with low density plasma. However, for the perturbations seen in Figs. \ref{av_temp_rho_time20_974} and \ref{av_temp_rho_time100_974}, we see a co-spatial increase (or decrease) in both the temperature and the density. As such, they highlight regions of adiabatic heating (or cooling). In S1, these perturbations form horizontally across the plane-of-the-sky (POS) and obscure the initial structures. In S2, on the other hand, they predominantly appear in regions where \(|x| \text{ \& } |y| > 8 \text{ Mm}\) and do not obscure the initial density and temperature configuration. This is due to the deformed nature of the wave front at later times in S2, implying that regions of compression (or rarefaction) associated with the wave do not necessarily align along the line-of-sight (LOS), resulting in weaker LOS density (and temperature) perturbations.

\section{Forward modelling} \label{forward_modelling_section}

\subsection{Emission lines and LOS}

To produce the synthetic emission data, the numerical results are forward modelled using the FoMo code \citep{VanDoorsselaereAntolin2016} which generates optically thin EUV and UV emission lines using the CHIANTI atomic database \citep{DereLandi1997, LandiYoung2013}.

We choose to model the Fe \rom{12} 193\AA\ and Fe \rom{16} 335\AA\ emission lines, as these provide good coverage of the temperature range in our simulations. The rest wavelengths are \(193.509\text{ \AA }\) and \(335.409 \text{ \AA }\), with peak formation temperatures of \(\text{log}(T) = 6.20\text{ }(\sim1.57\text{ MK})\) and \(\text{log}(T) = 6.42\text{ }(\sim2.65\text{ MK})\), respectively. 

We examine four different LOS angles which are illustrated in Fig. \ref{los_angles} and Table \ref{los_angle_table}. Two of the angles lie in the \(xy\text{-plane}\) and cut across the magnetic flux tubes (LOS1 and 2) whereas LOS3 and LOS4 lie in the \(xz\text{-plane}\) and are approximately aligned with the magnetic structure. The LOS3 viewing angle is directly aligned with the axis of the loop, which could occur near disk centre in the lower atmosphere.

\begin{figure*}[t!]
  \centering
  \includegraphics[width=0.7\textwidth]{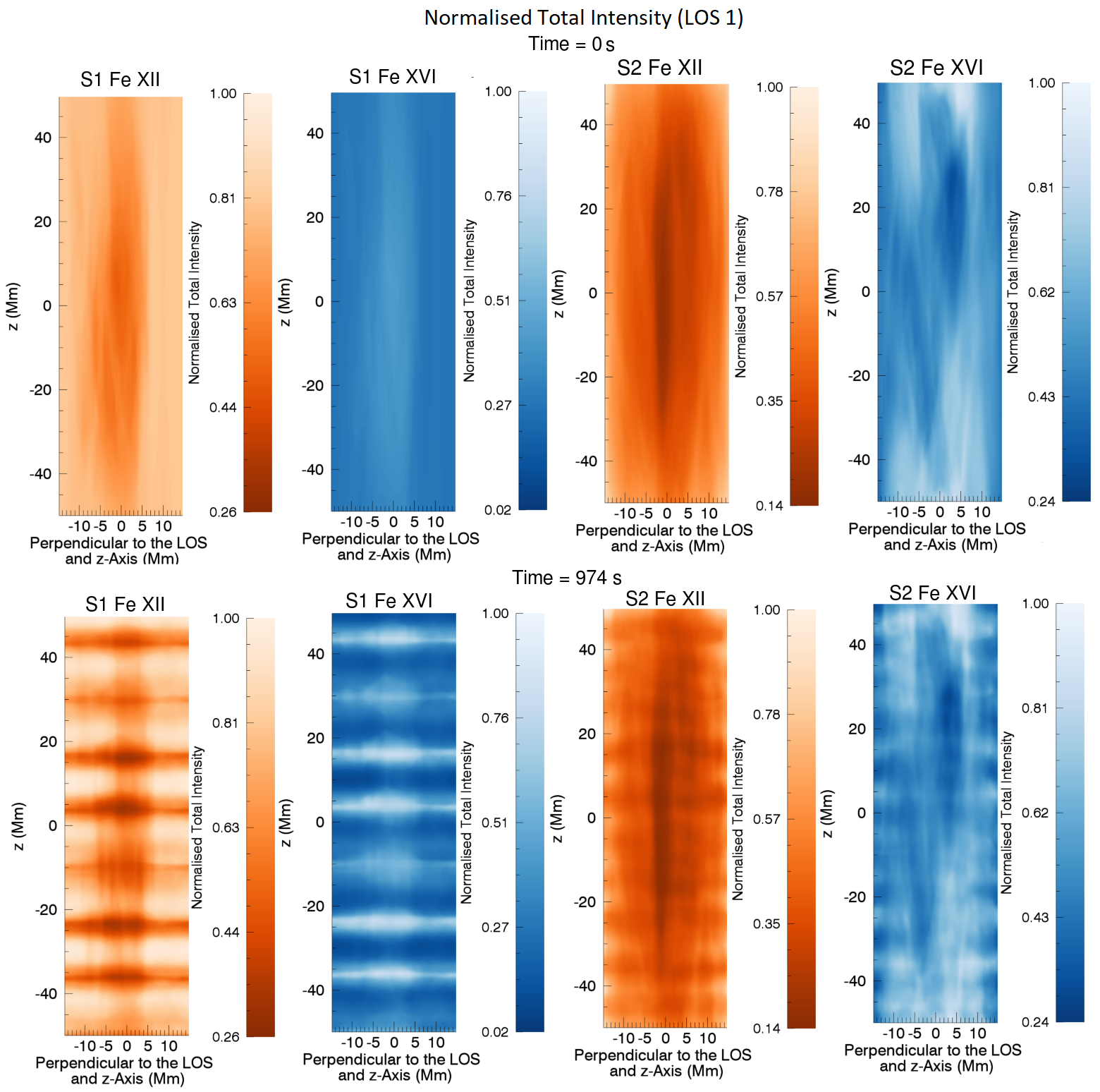}
  \caption{Normalised total intensity for S1 and S2 in Fe \rom{12} and Fe \rom{16} looking along LOS1 at 0s (row 1) and 974s (row 2). In each case, the intensities have been normalised by their own spatio-temporal maximum.}
  \label{tot_int_los1}
\end{figure*}

The numerical grid used in \citet{HowsonDeMoortel2020} was modified in this forward modelling analysis to reduce computational costs by spatially resampling the original simulations to every fourth grid cell along \(x\), \(y\) and \(z\). The temporal resolution was not altered. From examining a selection of the data at full spatial resolution, we confirmed that the spatial re-sampling had no significant impact on the forward modelling results.

\subsection{Imaging signatures}

\begin{figure*}[htpb!]
  \centering
  \includegraphics[width=0.7\textwidth]{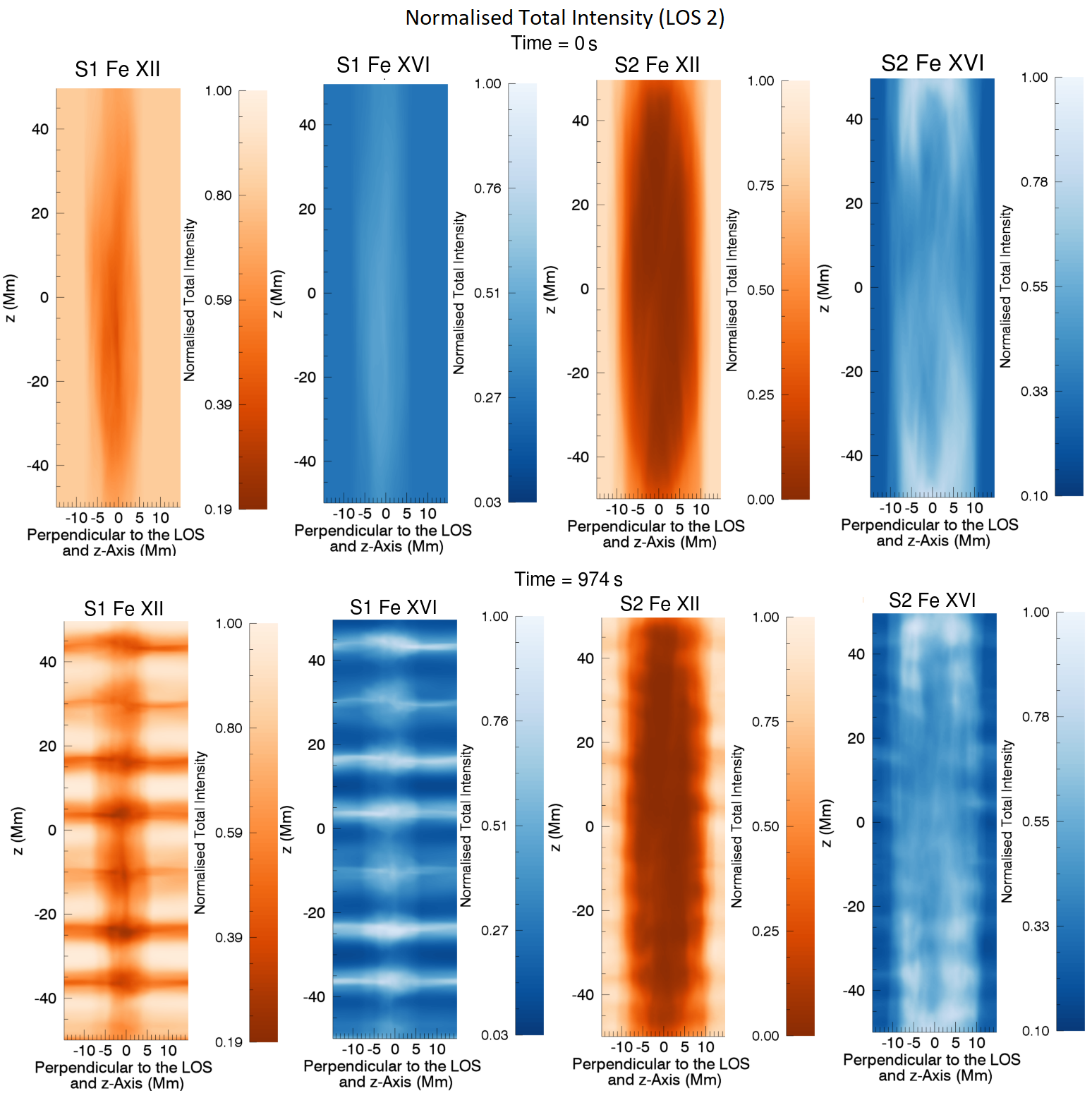}
  \caption{Normalised total intensity for S1 and S2 in Fe \rom{12} and Fe \rom{16} looking along LOS2 at 0s (row 1) and 974s (row 2). In each case, the intensities have been normalised by their own spatio-temporal maximum.}
  \label{tot_int_los2}
\end{figure*}

Figures \ref{tot_int_los1} and \ref{tot_int_los2} illustrate the synthetic total intensity at two instances in time (t = 0 s and t = 974 s) in Fe \rom{12} and Fe \rom{16}. We show results for both simulations, along LOS1 (Fig. \ref{tot_int_los1}) and LOS2 (Fig. \ref{tot_int_los2}).  Fe \rom{16} generally detects the hotter plasma i.e.~initially the central column and later on, the regions where we observed adiabatic heating. Of course, if the plasma density is low, then even in regions of hot plasma, the Fe \rom{16} intensity will also be low. An example of this can be seen in the S2 simulation (last column of Fig. \ref{tot_int_los1}; \(|x| \lesssim 7\) Mm, \({-35 \text{ Mm} \lesssim y \lesssim 40 \text{ Mm}}\)). The low intensity in the cooler line, for S2, is a combination of low density and plasma being outwith the formation temperature of Fe \rom{12} (third column of Figs. \ref{tot_int_los1} \& \ref{tot_int_los2}). Apart from this, the Fe \rom{12} intensity is essentially the `inverse' of Fe \rom{16}.

 We note that the complexity in the equilibrium configuration is not evident from the intensity contours at $t=0$ \citep[see, for example,][]{PontinJanvier2017}. The hotter line for simulation S1 is the only case where a flux tube-like structure is even detected. As expected, there are no signs of the two threads in S1 which became unstable during the \citet{ReidHood2018} simulation.  However, there are also no signs of the third thread, which is still distinguishable along LOS1 when examining the traced magnetic field lines in Fig. \ref{los_angles}.

 \begin{figure*}[htpb!]
\centering
\includegraphics[width=0.75\textwidth]{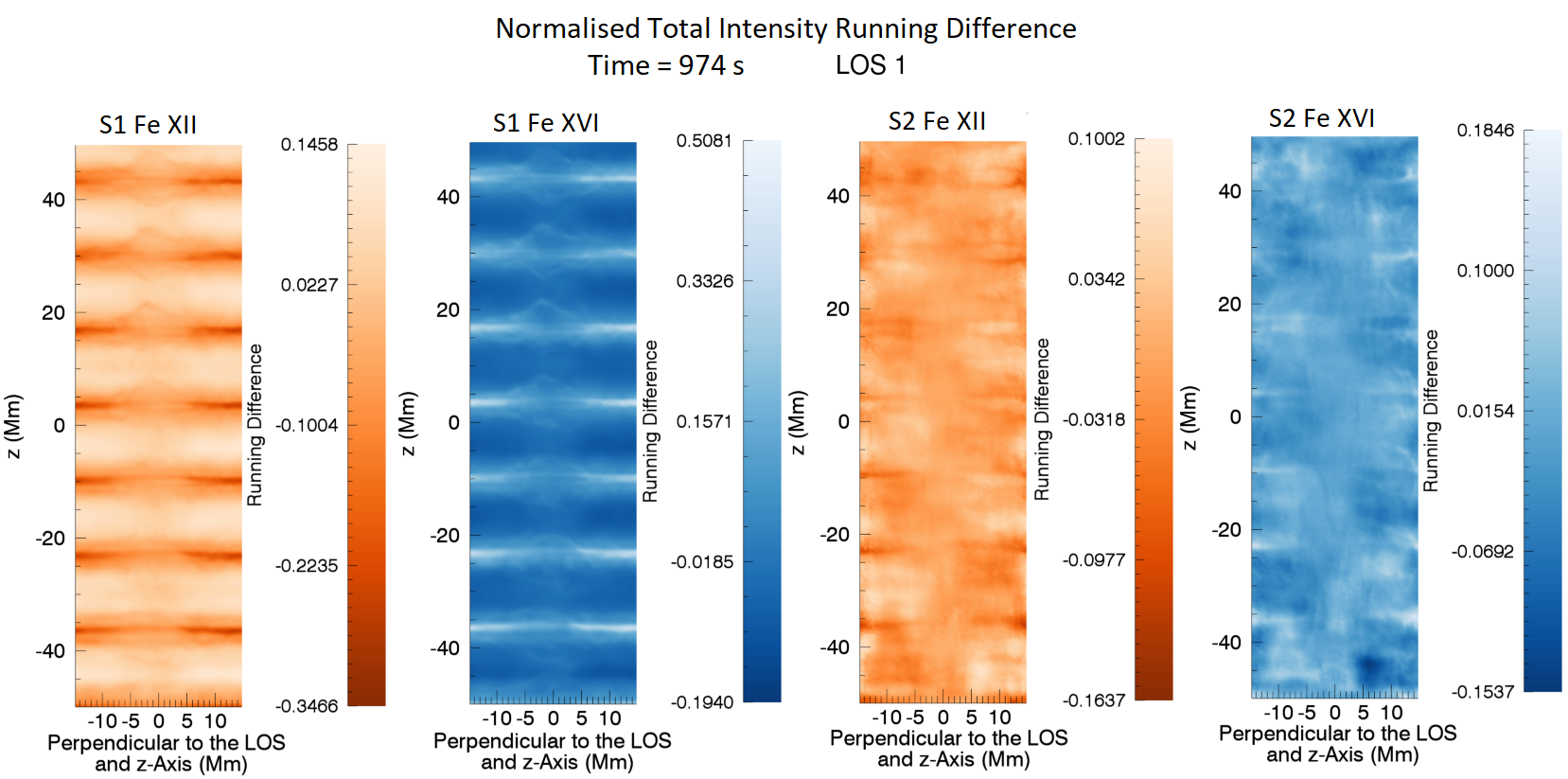}
\caption{Normalised running difference of total intensity (with a cadence of 3.6 s) for S1 and S2 in Fe \rom{12} and Fe \rom{16} looking along LOS1 at 974s. In each case, the running-differences have been normalised by their own spatio-temporal maximum.}
\label{rundiff_974_los1}
\end{figure*}

\begin{figure}[t!]
\centering
\vspace{0cm}
\begin{subfigure}{0.45\textwidth}
  \centering
  \hspace{0cm}
  \makebox[0pt]{\includegraphics[width=1.\textwidth]{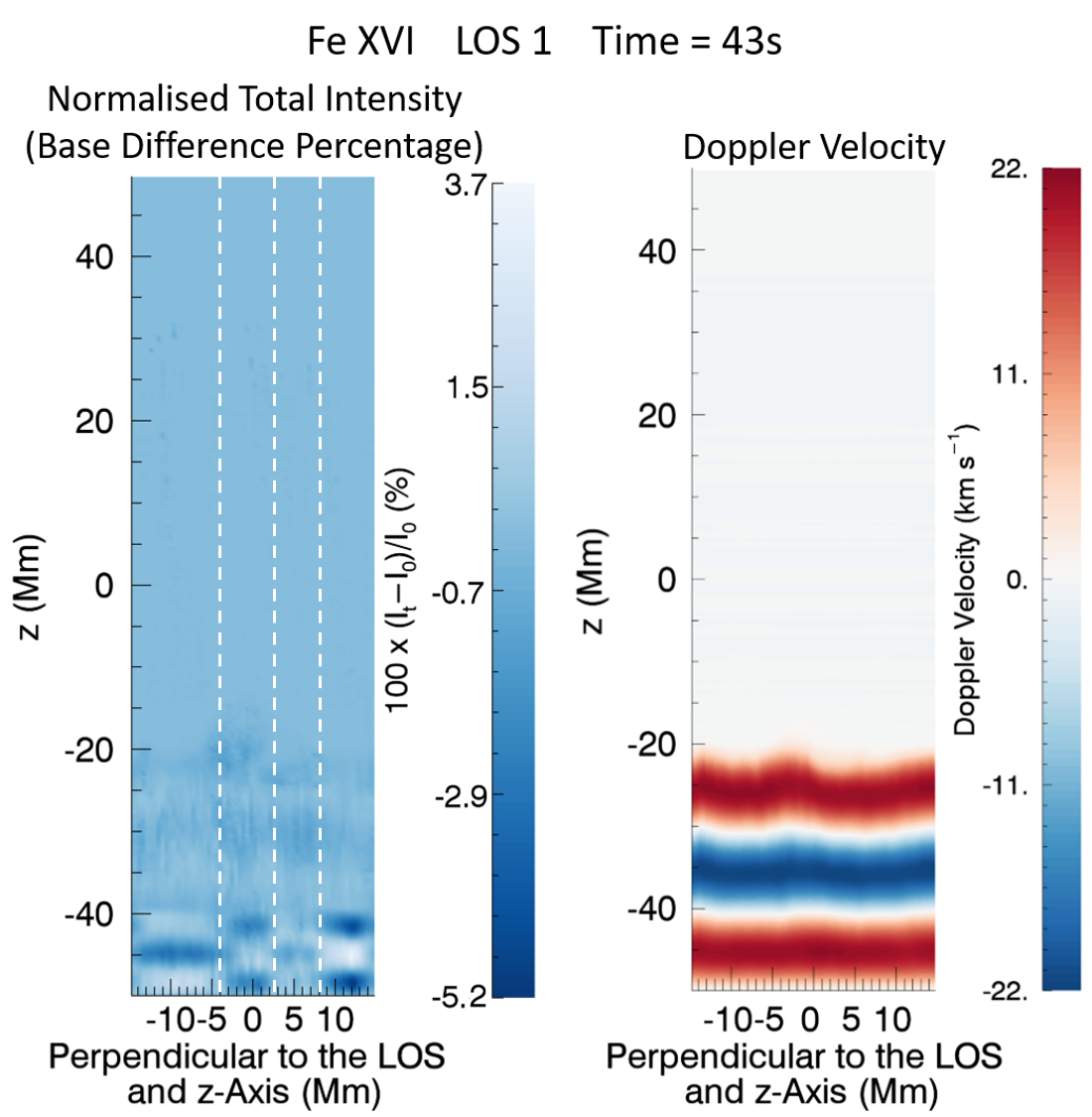}}
  \caption{}
  \label{basediff_int_and_dv}
\end{subfigure}
\begin{subfigure}{0.35\textwidth}
\hspace{0cm}
  \centering
  \hspace{0cm}
\makebox[0pt]{\includegraphics[width=1.\textwidth]{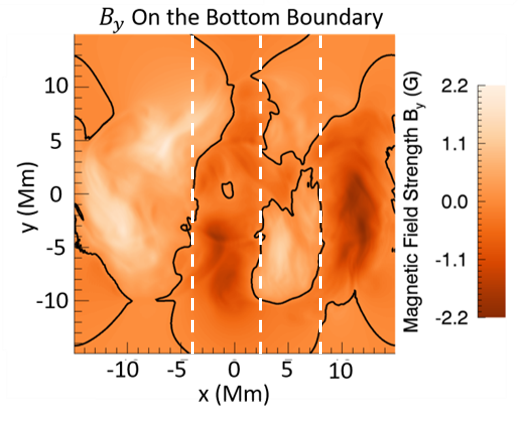}}
  \caption{}
  \label{By_bottom_boundary}
\end{subfigure}
\caption{First and second panel of (a) illustrate the total relative intensity base difference and Doppler velocity, respectively, in Fe \rom{16} along LOS1 in S2. Depicted in (b) is \(B_y\) on the bottom \(z\) boundary. The black line represents the \(B_y = 0\) G contour. In (a) \& (b) the white dashed lines highlight the location of the small horizontal spatial features. All panels are at 43 s.}
\end{figure}

 \begin{figure}[t!]
\centering
\includegraphics[width=0.4\textwidth]{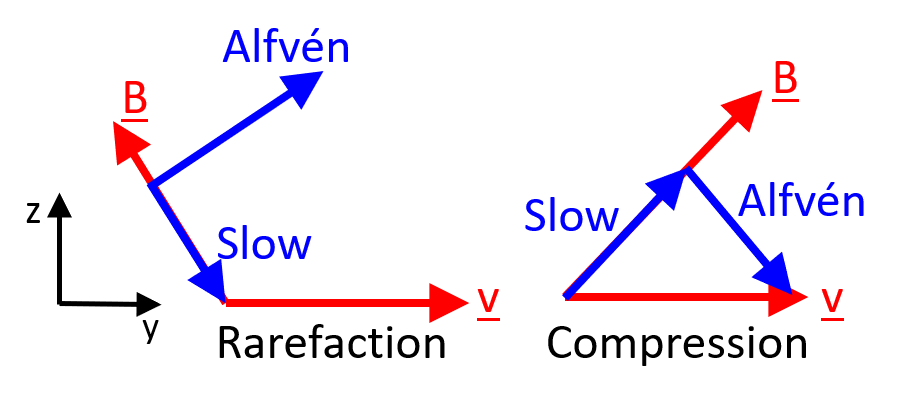}
\caption{Illustration of the compression and rarefaction due to the slow wave, seen in the intensity base difference in Fig. \ref{basediff_int_and_dv}. All vectors are projected onto the $yz$-plane, where $\textbf{B}$ is the magnetic field on the lower boundary, $\textbf{v}= (0,v_y,0)$ is the boundary driver and \Alfven and Slow are the \Alfven and slow components of the propagating wave.}
\label{basediff_43s_explained}
\end{figure}

 Due to the compressible nature of the waves, we do expect to see some intensity changes, associated with the density and temperature perturbations discussed above. Indeed, as the simulations progress, the presence of waves becomes most apparent for the least complex magnetic field configuration (see the first and second panel on the bottom row of Figs. \ref{tot_int_los1} and \ref{tot_int_los2}), where we saw the clearest density and temperature perturbations. For the more complex field (third and fourth panels), the wave-like behaviour is less apparent and could be mistaken for background noise. Figure \ref{rundiff_974_los1}, which shows the running difference of the total intensity (i.e. the difference between subsequent time steps with a cadence of 3.6 s), has the same properties. We see a clear signature of the waves in both lines for S1 (first two panels). This change in intensity highlights the compressible nature of the velocity field despite the incompressible nature of the boundary driver. However, for the more braided field (S2 - last two panels), the running difference does not reveal the wave dynamics either, due to the complex, continuously changing nature of the wave front.

In addition to the running difference, we investigate the base difference i.e. we subtract the initial intensity from the intensity at each time step. This is shown in the first panel of Fig. \ref{basediff_int_and_dv}. We observe minor changes in the intensity profile near the start of the simulations as illustrated at 43 s during S2 (first panel). These changes coincide with the front of the \Alfvenic wave (at \(z \approx -20 \text{ Mm}\)) seen in the Doppler velocity in the second panel (a more in-depth analysis of the Doppler velocity is given in Sect. \ref{spec_sig}). The presence of a slow wave, generated by the boundary driver,  can also be seen in the left-hand panel, propagating behind the transverse wave with a wave front at \(z \approx -40 \text{ Mm}\). These perturbations to the intensity are caused by the compression and rarefaction of the plasma as a result of the waves. Intriguingly, the horizontal spatial scales of the slow wave (seen in the base difference), are smaller than those seen in the transverse wave (shown by the Doppler velocity). This is due to the way the boundary driver (\(v_y\)) interacts with the $y$-component of the magnetic field on the bottom boundary. The schematics in Fig. \ref{basediff_43s_explained} helps to explain why this is the case. When \(B_y v_y>0\), the slow wave component (`Slow' in Fig. \ref{basediff_43s_explained}) acts along the magnetic field, causing compression, whereas when \(B_y v_y<0\),  the slow component is now anti-parallel to the magnetic field, resulting in rarefaction of the plasma. In Fig. \ref{By_bottom_boundary}, we show the configuration of \(B_y\) on the bottom boundary of the domain. We see that the sign of $B_y$ broadly aligns with the small spatial scale features seen in the base difference and matches the compression and rarefaction pattern illustrated in Fig. \ref{basediff_43s_explained}.

\begin{figure*}[t!]
\centering
\vspace{0cm}
\begin{subfigure}{0.48\textwidth}
  \centering
  \hspace{0cm}
  \makebox[0pt]{\includegraphics[width=1.\textwidth]{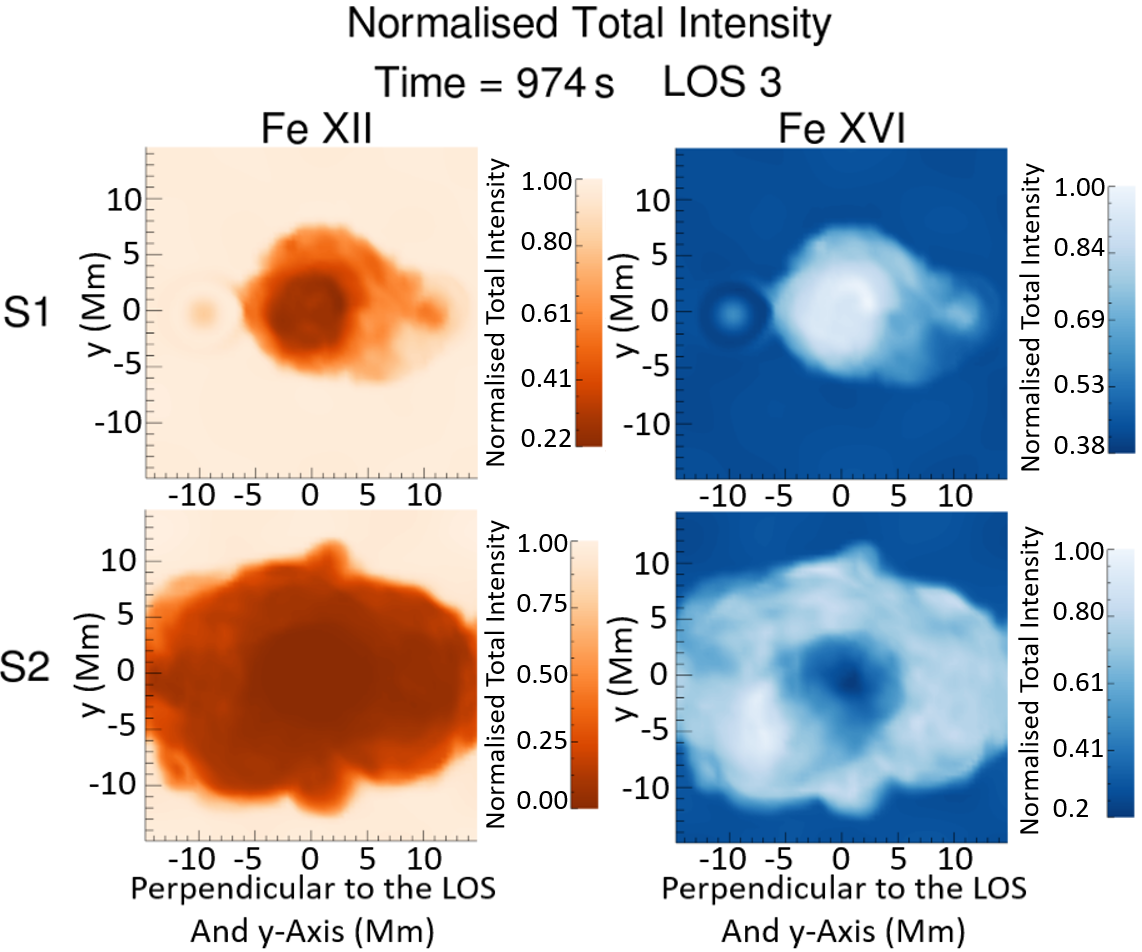}}
  \caption{}
  \label{int_los3_time100_974}
\end{subfigure}%
\begin{subfigure}{0.5\textwidth}
\hspace{0cm}
  \centering
  \hspace{0cm}
\makebox[0pt]{\includegraphics[width=1.\textwidth]{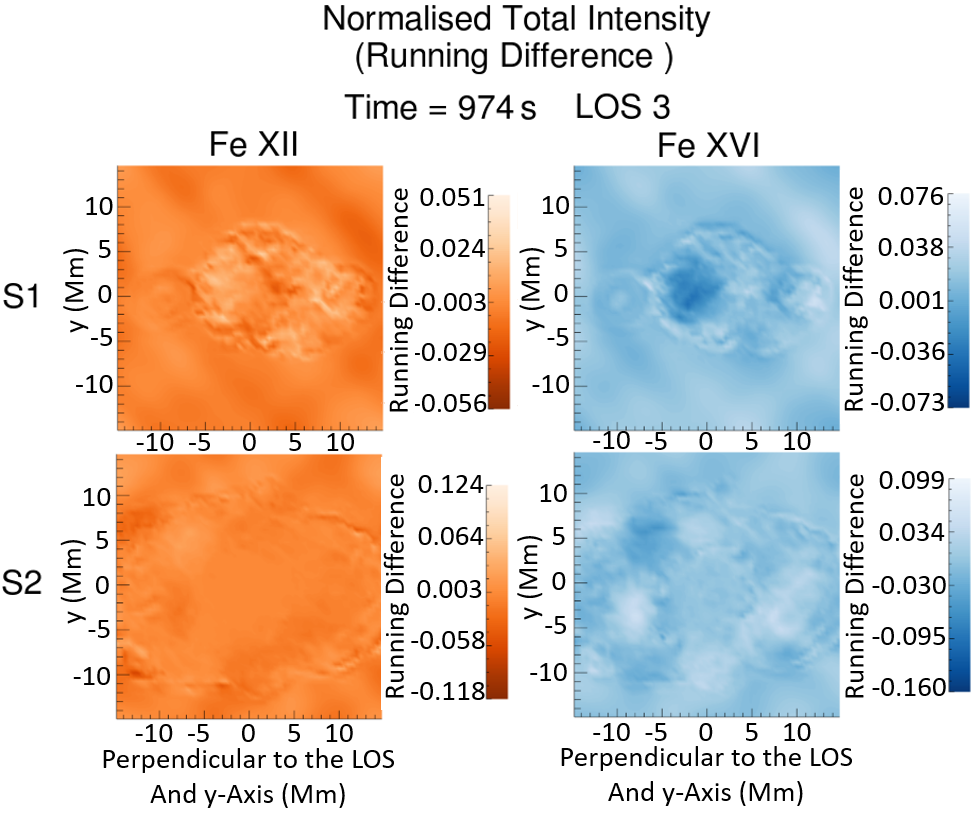}}
  \caption{}
  \label{rundiff_bottom_974}
\end{subfigure}
\caption{Normalised (a) total intensity and (b) running difference (with a cadence of 3.6s) along LOS3 in Fe \rom{12} (first column) and Fe \rom{16} (second column) in both S1 (first row) and S2 (second row) at 974s. Quantities in each panel have been normalised by their own spatio-temporal maximum.}
\label{int_rundiff_los3}
\end{figure*}

The figures of intensity and running difference along LOS1 and 2 (Figs. \ref{tot_int_los1}, \ref{tot_int_los2} \& \ref{rundiff_974_los1}), do not show obvious signs of magnetic field complexity. This also holds for the intensity along LOS3 (Fig. \ref{int_los3_time100_974}). However, when examining the running difference along LOS3 (Fig. \ref{rundiff_bottom_974}), the small scale spatial structuring becomes apparent. This helps identify regions of complex field but does not imply that other regions are not complex. For example, \(\text{Fe \rom{12}}\) S2 (Fig. \ref{rundiff_bottom_974} - bottom left panel) does not show small scale structures in the centre, despite complex field being present (compare to bottom left panel of Fig. \ref{los_angles}). Hence, if observing the Fe \rom{12} emission in isolation, one could incorrectly conclude that the S1 magnetic field is more intricate. Similarly, when examining the Fe \rom{16} synthetic emission in isolation (Fig. \ref{rundiff_bottom_974} - right column), even though it shows the regions of complex magnetic field projected onto the plane of the sky (compare to left column of Fig. \ref{los_angles}), distinguishing the intricacy of the two magnetic fields (S1 and S2) is not straightforward. Therefore, a comparison of the complexity of the two magnetic field configurations from the intensity images does not necessarily give the correct results.

\begin{figure*}[ht!]
\centering
\includegraphics[width=0.75\textwidth]{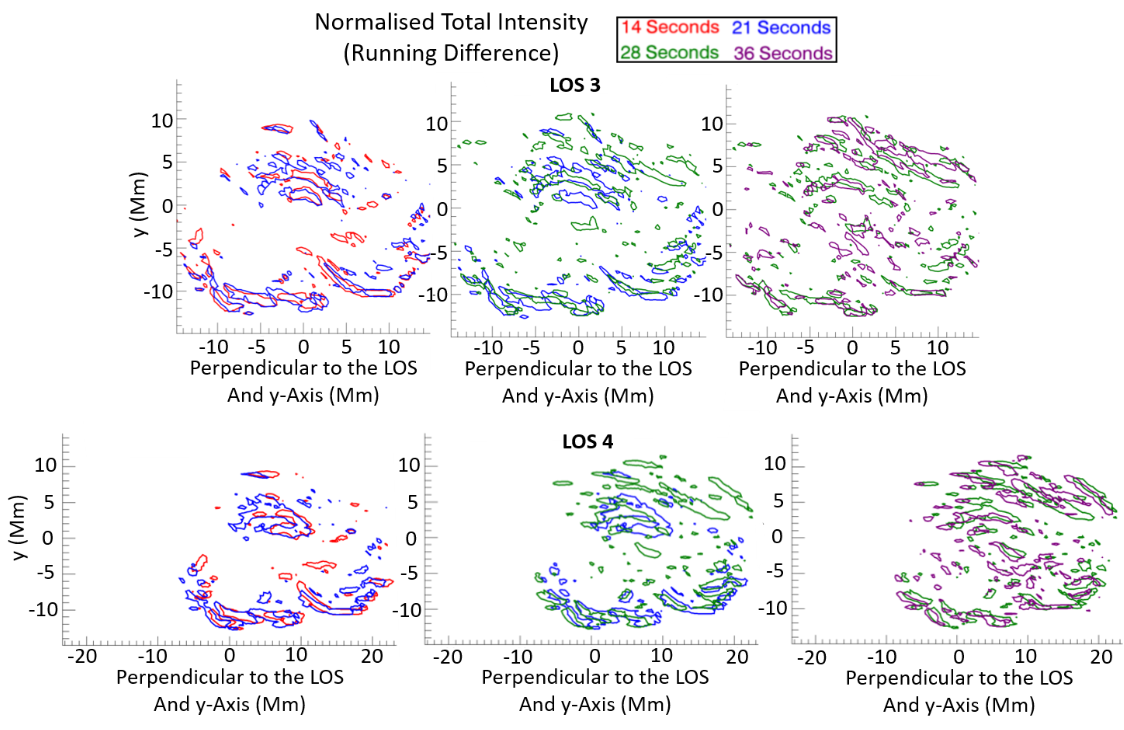}
\caption{Single contour level of the normalised running difference of the total intensity (with a cadence of 3.6s) along LOS3 (row 1) and LOS4 (row 2) in Fe \rom{16} during S2 at 14 (red), 21 (blue), 28 (green) and 36 (purple) seconds.}
\label{rundiff_los3_and_4_time100_fe16_contourlines}
\end{figure*}

One intriguing feature visible in the running difference of intensity along LOS3 \& 4 during the first transit of the transverse waves through the domain, is the presence of an apparent rotational motion (see movie1). This is not indicative of real torsional motions, which are not present within the plasma but instead is caused by compressible wave fronts propagating along twisted field. In Fig. \ref{rundiff_los3_and_4_time100_fe16_contourlines}, we have illustrated this behaviour by tracking one contour level of the running difference at consecutive times during S2 (it is also seen in S1). Along LOS3, we see an apparent anticlockwise rotation but by altering the viewing angle by just \ang{10} (LOS4), the rotational motion is hardly observable. For tilted viewing angles, the upward propagation of waves from the lower boundary (a motion from right to left in the left panel of movie1 and the bottom row of Fig. \ref{rundiff_los3_and_4_time100_fe16_contourlines}) obscures the apparent rotation.

In addition, this apparent torsional motion is only visible during the initial transit of the waves through the domain, i.e. before the first reflection off the top boundary. Once the waves reflect, the interference between the counter-propagating waves leads to random and chaotic patterns. Hence, this apparent rotational motion is probably unlikely to be observed, for example in closed coronal loops, due to the constant motions and waves propagating along magnetic field lines causing wave interference. However, in cases where there is a significant difference in the outward and inward propagating wave power \citep[e.g.][]{VerthTerradas2010, TiwariMorton2019} or in open magnetic field regions \citep[e.g.][]{MortonTomczyk2015}, this apparent rotational motion may lead to misinterpretations of torsional wave modes.

\subsection{Spectral signatures} \label{spec_sig}

\begin{figure*}[t]
\centering
\includegraphics[width=0.7\textwidth]{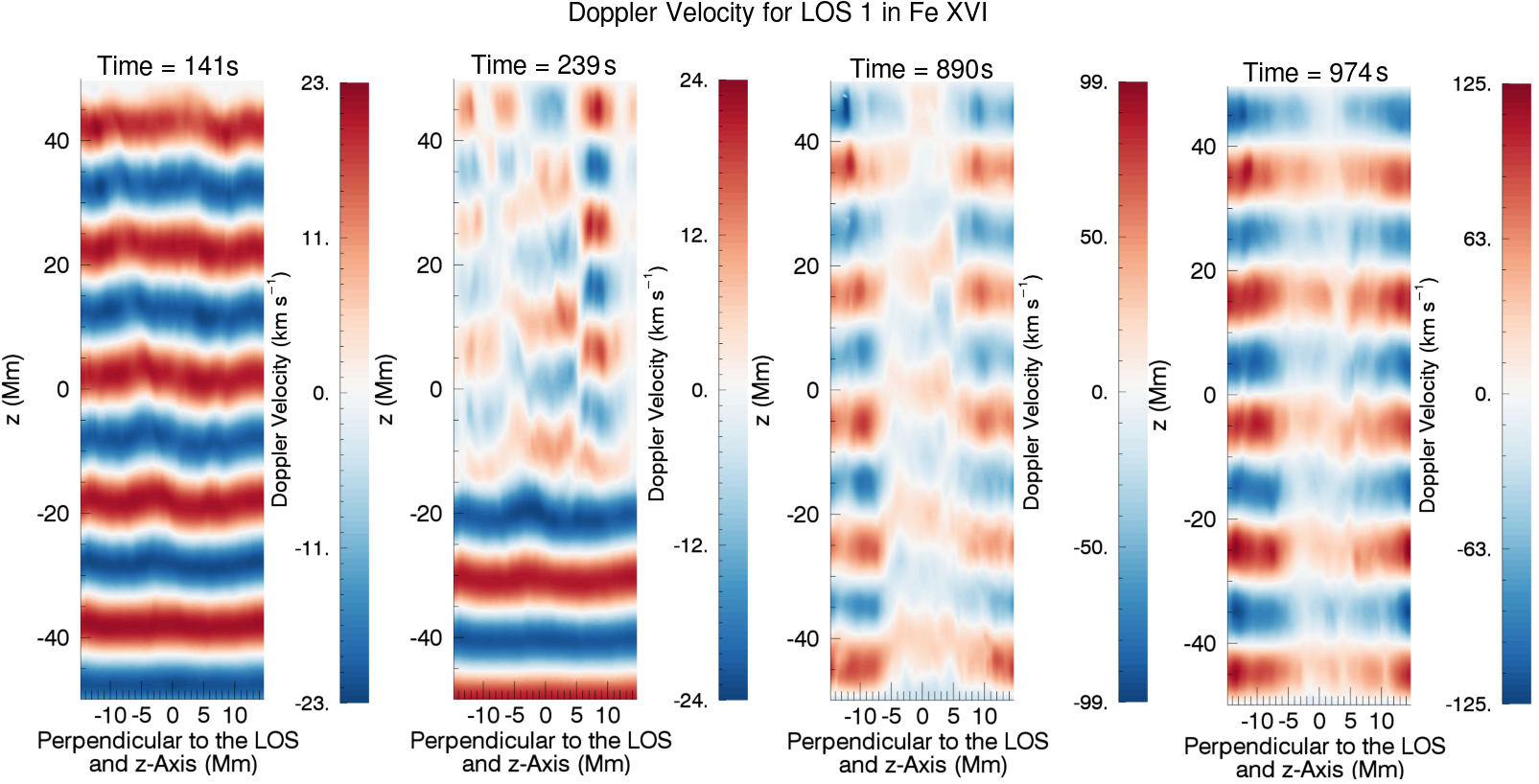}
\caption{Doppler velocity for S2 along LOS1 in Fe \rom{16}. Note that the range on the colour bar changes in each panel.}
\label{dv_time100_fe16_los1}
\end{figure*}

\begin{figure}[ht]
\centering
\makebox[0pt]{\includegraphics[width=0.22\textwidth]{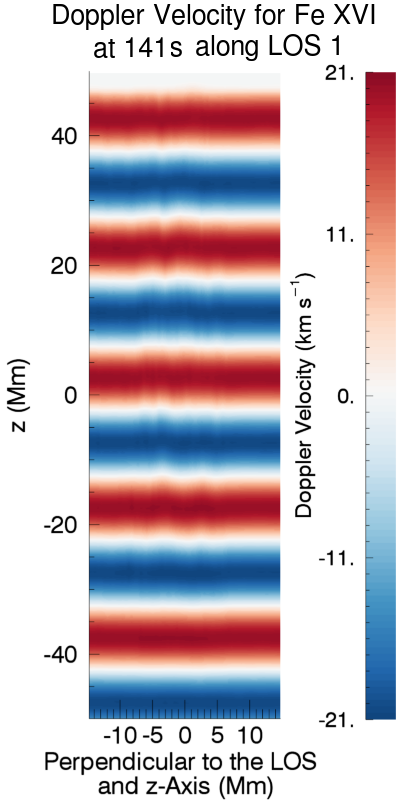}}
\caption{Doppler velocity from S1 in Fe \rom{16} at 141 s along LOS1.}
\label{dv_time20_los1_fe16_141}
\end{figure}

\begin{figure}[ht]
\centering
\makebox[0pt]{\includegraphics[width=0.22\textwidth]{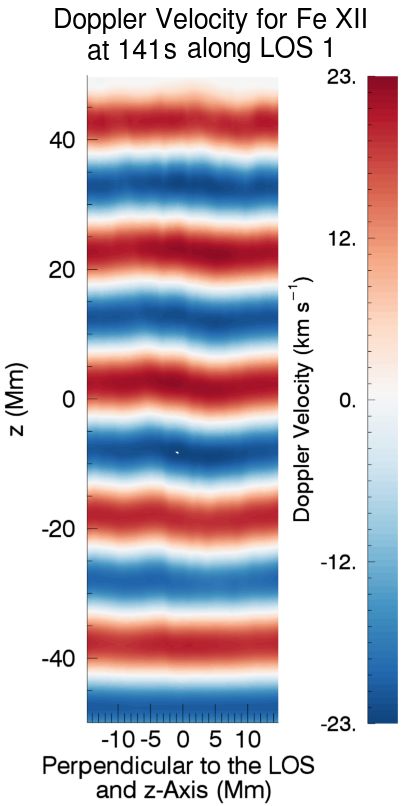}}
\caption{Doppler velocity from S2 in Fe \rom{12} at 141 s along LOS1.}
\label{dv_time100_los1_fe12_141}
\end{figure}

Even though there is a weak compressible aspect to the waves, they do not show strong signatures in intensity. In addition, increased LOS cancellation for the more braided simulation leads to very weak, if any, detectable signs of wave propagation in the S2 simulation (last two panels in lower rows of Figs. \ref{tot_int_los1} \& \ref{tot_int_los2} and final two panels in Fig. \ref{rundiff_974_los1}).

However, we are able to clearly detect the presence of waves using Doppler velocities. In Fig. \ref{dv_time100_fe16_los1}, we illustrate the Doppler velocities observed with the \({\text{Fe \rom{16}}}\) line along LOS1. These are obtained by fitting a (single) Gaussian to the specific intensity and measuring the shift of the Gaussian peak from the rest wavelength. The times in the first two panels coincide with Fig. \ref{tom_vy_s2} and the remaining two panels show later times in the simulation, when the wave interference and phase mixing are well-developed. At all times, propagating waves are clearly present as perturbations in the Doppler velocities. The magnitudes of the Doppler shifts are somewhat large in comparison to actual observations. However, when an integration time of 10 s (similar to that of modern spectrometers) was used, we found a decrease in the values of the Doppler shift by a factor of almost two. The structure of the Doppler velocity profiles was largely unaltered. 

We note that in S2, a single Gaussian could not be fitted to a very small percentage (less than $\sim$0.06\%) of the line profiles. These instances all occurred at later times in the simulation when multiple, more complex plasma flows are present along the LOS, resulting in complex, double (or more) peaked line profiles. As this only happens in a very small number of cases, it does not affect the analysis of the data presented here.

\begin{figure}[ht!]
\centering
\includegraphics[width=0.35\textwidth]{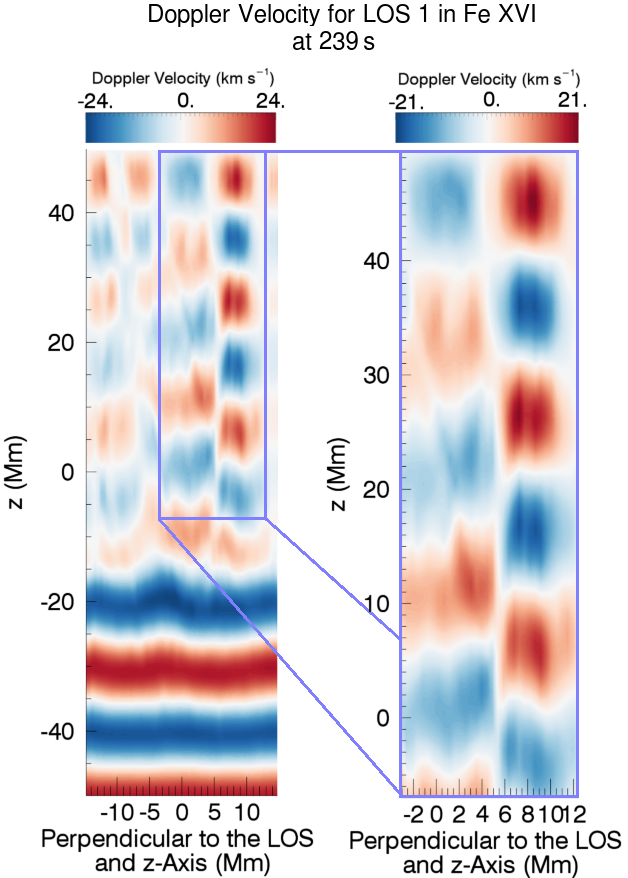}
\caption{Doppler velocity for S2 along LOS1 in Fe \rom{16} (left panel), repeated from panel 2 of Figure \ref{dv_time100_fe16_los1}. A zoomed in region of the domain (right panel) where \({-7 \text{ Mm }\leq z \leq 50 \text{ Mm}}\) and the other axis is between \(-3 \text{ Mm}\) and \(12\text{ Mm}\). Note the change in the range of the colour bars.}
\label{dv_time100_fe16_239_section}
\end{figure}

Before the transverse waves reflect off the top boundary (\(t = 141 \) s), we see (weak) phase mixing as described in \citet{HowsonDeMoortel2020} and in Sect. \ref{evolution}. The level of complexity is evident by contrasting S2 with S1 using the same emission line (compare the left hand panel of Fig. \ref{dv_time100_fe16_los1} and Fig. \ref{dv_time20_los1_fe16_141}). As expected, the level of distortion to the wave front is greater in S2, confirming that the more complex field leads to more phase mixing. When comparing the S2 Doppler velocities in Fe \rom{12} (Fig. \ref{dv_time100_los1_fe12_141}) and in Fe \rom{16} (first panel Fig. \ref{dv_time100_fe16_los1}), there is more phase mixing evident in the hotter line. This is due to the temperature profile of the background magnetic field. In general, the more complex field (which enhances phase mixing) has higher local temperatures. The cooler lines, on the other hand, mostly detect regions where the field is less complex, and thus, exhibit less wave distortion.

\begin{figure*}[t!]
\centering
\includegraphics[width=0.7\textwidth]{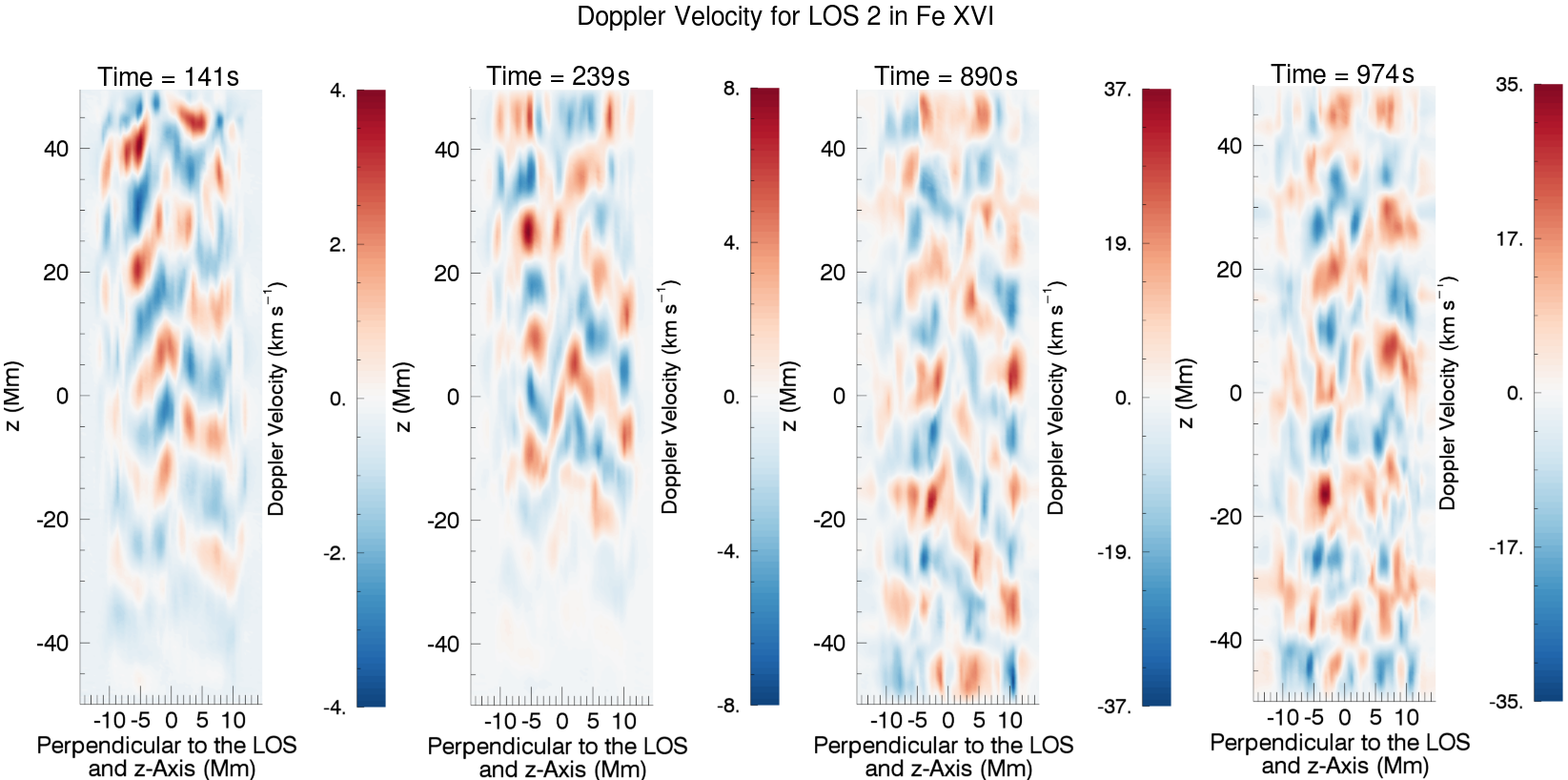}
\caption{Doppler velocity for S2 along LOS2 in Fe \rom{16}. Note the range on the colour bar changes in each panel.}
\label{dv_time100_fe16_los2}
\end{figure*}

Looking more closely at a small region of panel two in \({\text{Fig. \ref{dv_time100_fe16_los1}}}\), as illustrated in Fig. \ref{dv_time100_fe16_239_section}, an out of phase red-blue structure is detected. Assuming no prior knowledge of this simulation (i.e.~the magnetic field configuration and the polarisation of the boundary driver) and a field of view limited to this small region, one could misinterpret this Doppler velocity configuration as torsional motions.

Similar Doppler velocity features have been identified by various authors in observational data. For example, \citet{SrivastavaShetye2017} find a similar signature when examining the Doppler velocity of a highly structured magnetic flux tube, using SST/CRISP in the \(H_\alpha\) 6562.8\(\text{ \AA }\) spectral line which was interpreted as torsional oscillations. \citet{KohutovaVerwichte2020} also interpret such alternating red-blue shifts along the edges of a flux tube as torsional \Alfven oscillations. However, in our simulations, these specific Doppler velocity patterns are the result of phase mixing and counter-propagation of transverse waves in a complex magnetic field rather than actual torsional waves.

Figure \ref{dv_time100_fe16_los2} illustrates the LOS2 Doppler velocities at the same times as the LOS1 Doppler velocities in Fig. \ref{dv_time100_fe16_los1}. The change in the wave polarisation, discussed in \citet{HowsonDeMoortel2019} and Sect. \ref{evolution}, is evident when comparing the Doppler shifts along the different LOS angles. Energy is transferred from the \(v_y\) component (observable in LOS1) of the velocity field near the bottom boundary to a \(v_x\) component (observable in LOS2) as time progresses. From the first panel, we see that as the waves propagate up through the domain, the change in polarisation becomes more pronounced, resulting in the Doppler velocities in LOS2 increasing with height.

\begin{figure}[ht]
\centering
\includegraphics[width=0.2\textwidth]{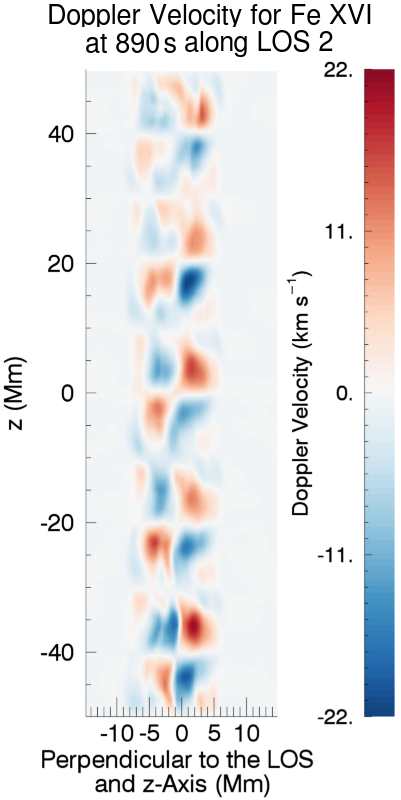}
\caption{Doppler velocities for Fe \rom{16} in S1 at 890s along LOS2.}
\label{dv_time20_los2_fe16_890}
\end{figure}

In LOS2, the spatial profile of the Doppler velocities in many locations is not dissimilar to the out of phase red-blue shifts seen in Fig. \ref{dv_time100_fe16_239_section}. Again, caution is needed to avoid misinterpreting these signatures of phase mixing as torsional motions. The LOS2 Doppler velocity signatures are quite complex and intricate, given that our boundary velocity driver is in fact a simple sinusoidal driver. Hence, complex observational signatures do not necessarily imply the presence of a complex driver but may instead (as in the case here) be indicative of a complex background magnetic field structure. When examining LOS2 in S1 (Fig. \ref{dv_time20_los2_fe16_890}), we find largely the same behaviour as in S2, except over a narrower region as less of the domain contains complex magnetic field structures (see top left hand panel of Fig. \ref{los_angles}). 

\begin{figure}[ht!]
\centering
\includegraphics[width=0.5\textwidth]{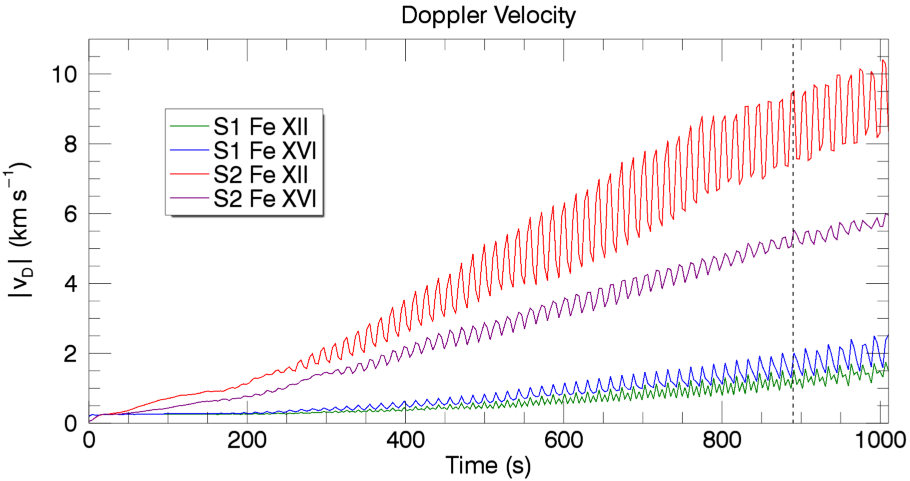}
\caption{Mean magnitude of the Doppler velocity along LOS2 from S1 (green and blue) and S2 (red and purple) in Fe \rom{12} (green and red) and Fe \rom{16} (blue and purple) as a function of time. The vertical black dashed line marks $t=890$ s corresponding to the second panel of Fig. \ref{dv_time100_fe16_los2} and Fig. \ref{dv_time20_los2_fe16_890}.}
\label{dv_los2_mean_plot}
\end{figure}

Figure \ref{dv_los2_mean_plot} shows the evolution of the mean magnitude of the Doppler velocity along LOS2 during both simulations, in both emission lines. It is clear that at all times and irrespective of the emission line, the values of the Doppler velocities are smaller for the S1 simulation. This is due to the less complex magnetic field structure in S1, which leads to a smaller transfer of energy to the $x$ component of the velocity field, (Fig. \ref{speed_vxvy_mean_plot}) and hence smaller Doppler velocity along LOS2. Figure \ref{speed_vxvy_mean_plot} shows the mean magnitudes of the $v_y$ and $v_x$ velocity components for both simulations. At early stages, the behaviour of $v_y$ is almost identical. Following the first reflection off the top boundary ($\sim 150 $ s), the oscillation amplitudes increase and we see $v_y$ (solid blue \& green lines) deviate in both simulations, as more energy is transferred to $v_x$ in S2 (dashed blue line) due to the more complex magnetic field.

\subsection{Kinetic energy budget}

\begin{figure*}[ht]
\centering
\begin{subfigure}{0.5\textwidth}
\centering
\makebox[0pt]{\includegraphics[width=1.\textwidth]{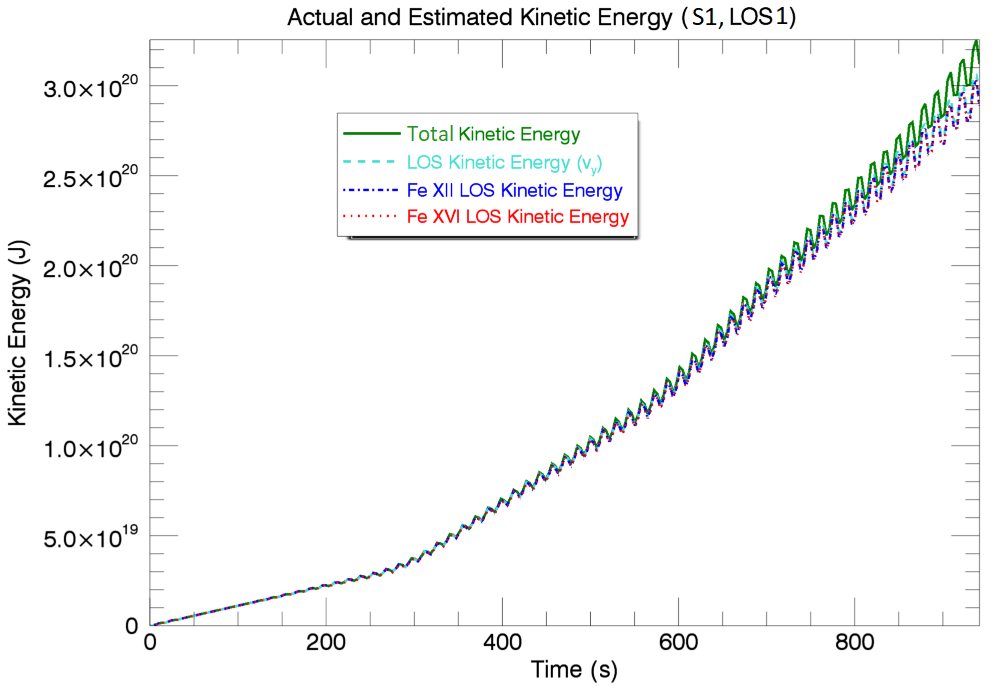}}
\end{subfigure}%
\begin{subfigure}{0.5\textwidth}
\centering
\makebox[0pt]{\includegraphics[width=1.\textwidth]{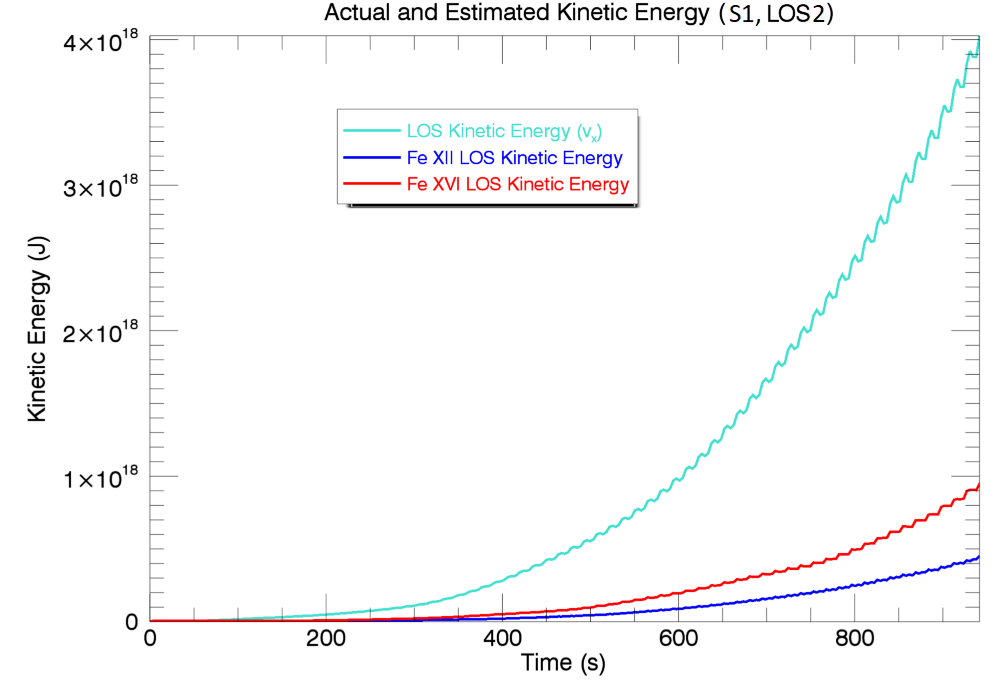}}
\end{subfigure}
\begin{subfigure}{0.5\textwidth}
\centering
\makebox[0pt]{\includegraphics[width=1.\textwidth]{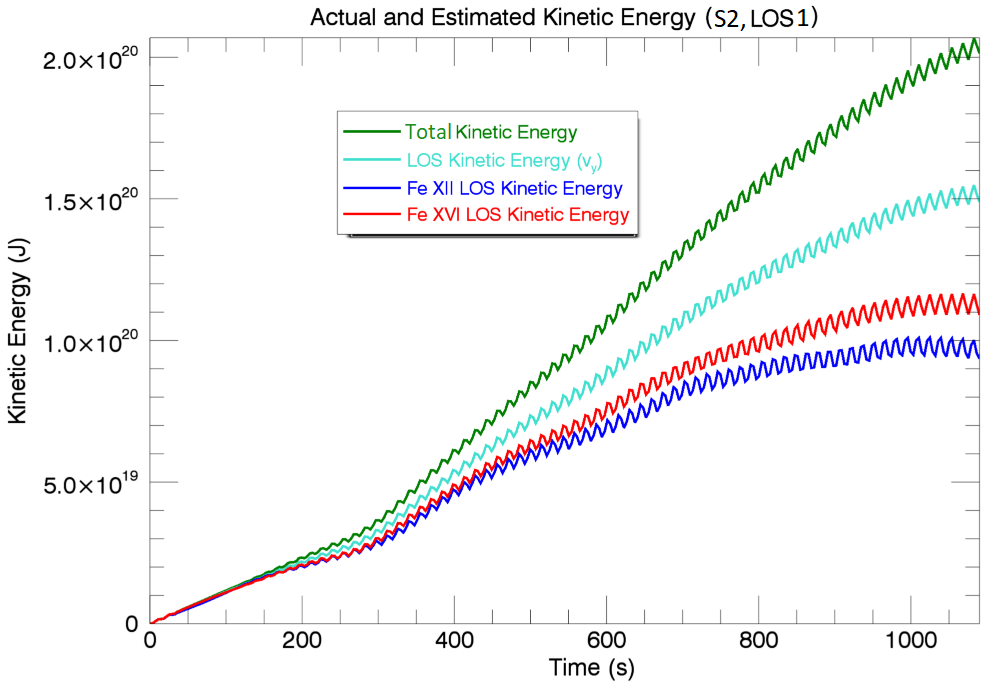}}
\end{subfigure}%
\begin{subfigure}{0.5\textwidth}
\centering
\makebox[0pt]{\includegraphics[width=1.\textwidth]{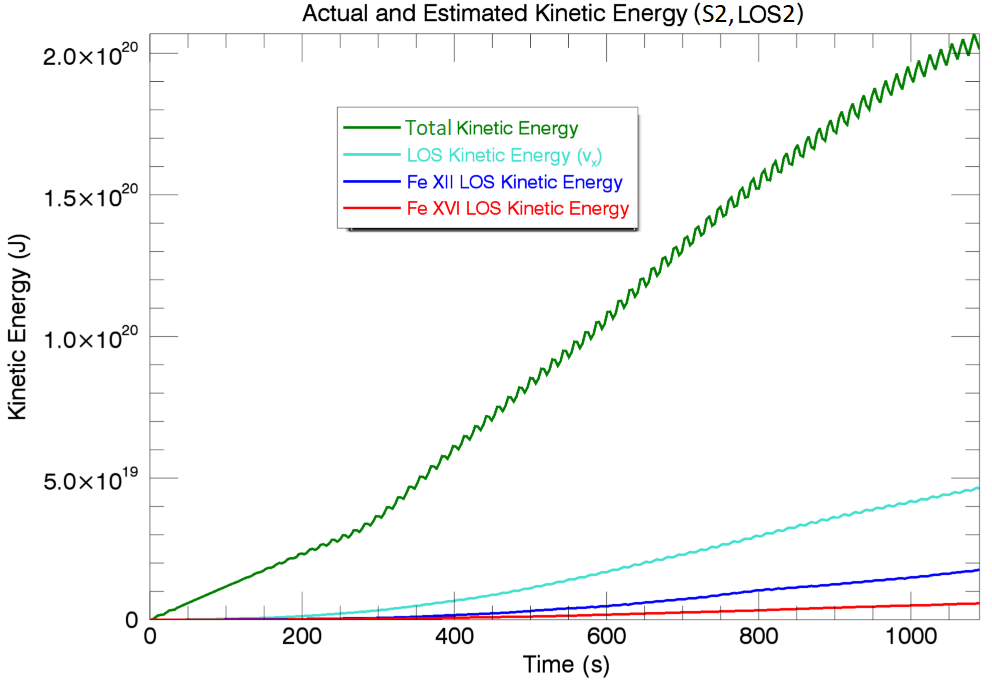}}
\end{subfigure}
\caption{Total (green) and LOS (turquoise) kinetic energy is integrated over the full 3D numerical domain compared to the estimated kinetic energy in Fe \rom{12} (blue) \& Fe \rom{16} (red), during S1 (row 1) \& S2 (row 2) along LOS1 (column 1) \& LOS2 (column 2). Note that all curves have been smoothed to illustrate the general trend rather than the amplitude of the oscillations. The total kinetic energy has not been plotted in panel (b) as it is several orders of magnitude larger (but the total kinetic energy is the same as in panel (a)).}
\label{ek_plots}
\end{figure*}

In this section, we compare the kinetic energy in both simulations to the estimated kinetic energy obtained from the synthetic emission data, i.e. using the Doppler velocities calculated in Sect. \ref{spec_sig}. Figure \ref{ek_plots} illustrates the evolution of these kinetic energies in both simulations (S1 \& S2), along LOS1 and 2. The volume integrated LOS kinetic energy (turquoise line) is calculated from the simulations results by setting $v_{LOS} = v_y$ or $v_x$, in \text{Eq. \ref{los_ke_eqn}}, for LOS1 and 2, respectively. 

\begin{eqnarray}
\;\text{LOS Kinetic Energy} &=& \frac{1}{2}\int\rho v_{LOS}^2 dV \label{los_ke_eqn} \\
\;\text{Estimated Kinetic Energy} &=& \frac{L}{2}\int\bar{\rho}v_D^2 dA \label{los_fomo_ke_eqn}  
\end{eqnarray}

To estimate the kinetic energies from the spectroscopic information (red and blue lines), we use Eq. \ref{los_fomo_ke_eqn}, where $L$ is the LOS depth, $\bar{\rho}$ is the average density and $v_D$ is the Doppler velocity for a given emission line, all along the LOS. Note that all the kinetic energies in Fig. \ref{ek_plots} have been smoothed to better illustrate their general evolution rather than the amplitude of the oscillations.

We realise that the depth, $L$, and density profile along the LOS are not easily measured from coronal observations. The length is simply a scaling factor which does not change the behaviour of the estimated kinetic energies. However, a reasonable estimate is required for comparison with the actual volume integrated kinetic energy. For the density, we require the average value along the line of sight. This can be estimated using line ratios as a density diagnostic tool \citep[e.g.][]{DereMason1979, MasonBhatia1979, LandiMiralles2014,PolitoDelZanna2016}. 

\begin{figure}[ht]
\centering
\includegraphics[width=0.5\textwidth]{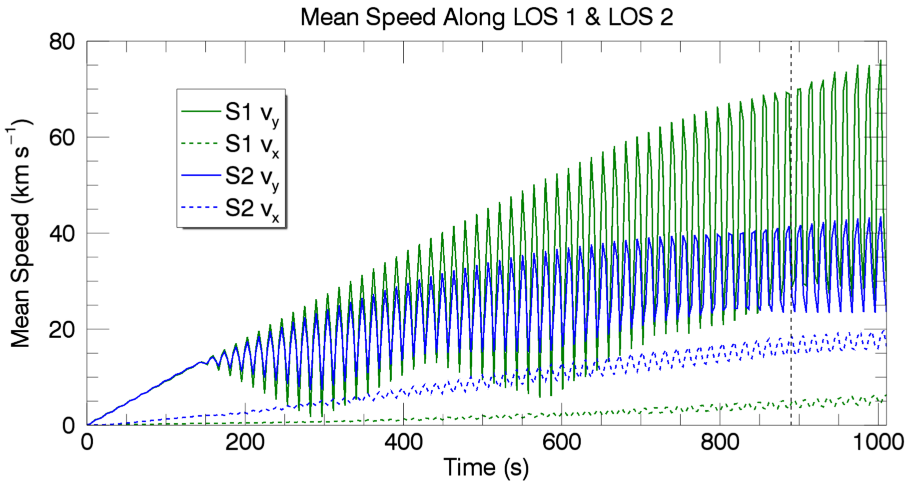}
\caption{Mean magnitude of $v_y$ (LOS1 solid lines) and $v_x$ (LOS2 dashed lines) in S1 (green) and S2 (blue) as a function of time. The vertical black dashed line marks $t=890$ s corresponding to the second panel of Fig. \ref{dv_time100_fe16_los2} and Fig. \ref{dv_time20_los2_fe16_890}.}
\label{speed_vxvy_mean_plot}
\end{figure}

Using a value of $L$ equal to 30 Mm (i.e.~the depth of our numerical domain), we can see that when we observe along the direction of the boundary driver (LOS1), the `observed' kinetic energy is a less accurate representation of the true (total) kinetic energy as the complexity of the field increases (i.e.~S2 is less accurate than S1 along LOS1 - see left column of Fig. \ref{ek_plots}). This is a result of the increased polarisation change in the S2 simulation where more energy is transferred from $v_y$ to $v_x$ (see Fig. \ref{speed_vxvy_mean_plot}) and hence we find a less accurate estimate of the total kinetic energy. Even so the estimates are, at worst, about 50\% of the total value.

Not only does the complexity of the magnetic field alter the estimated kinetic energy but the LOS angle has a significant effect as well. Comparing the left-hand and right-hand columns of Fig. \ref{ek_plots}, we see that for the LOS perpendicular to the boundary driver (i.e.~LOS2) the estimated kinetic energies are up to several orders of magnitude less than the total kinetic energy (compare blue line in top right panel with green line in top left panel of Fig. \ref{ek_plots}). As time progresses, and the polarisation of the wave changes, we observe an increase in the estimates along LOS2 and the estimated kinetic energies along LOS1 begin to deviate from the true total kinetic energy.

\begin{figure}[t!]
\centering
\makebox[0pt]{\includegraphics[width=0.45\textwidth]{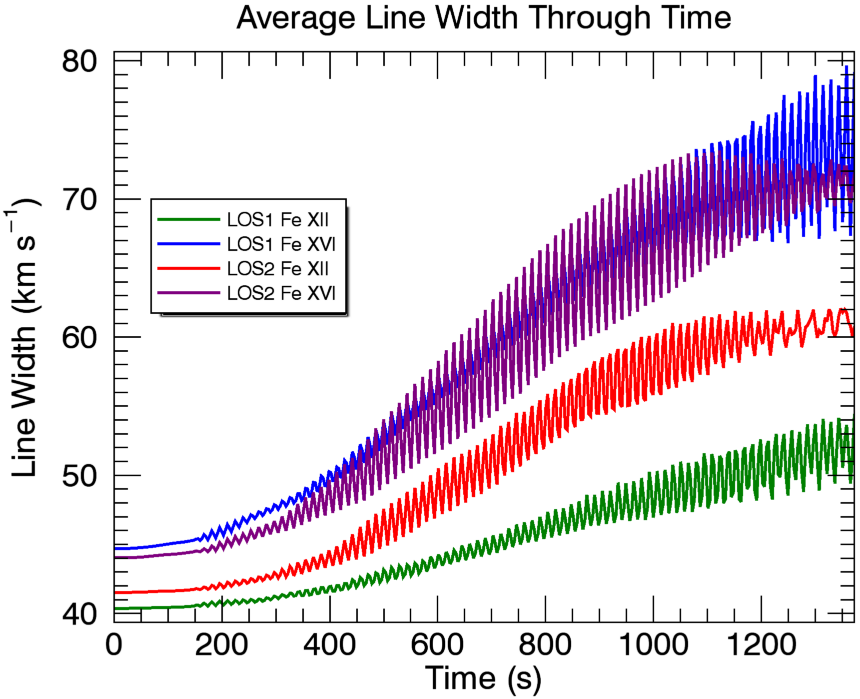}}
\caption{Average line width, using the FWHM, as a function of time during S2. We show LOS1 (green and blue) and LOS2 (red and purple) in Fe \rom{12} (green and red) and Fe \rom{16} (blue and purple).}
\label{av_lw_s2}
\end{figure}

As well as underestimating the total kinetic energy in the 3D domain, the estimated kinetic energy also underestimates the LOS kinetic energy. This is due to a combination of only sampling regions which are within the formation temperature range for the selected emission line and cancellation of velocities along the LOS. The energy which is `lost' due to multiple plasma flows along the LOS, caused by the complex magnetic field, would result in an increased non-thermal line width of the specific intensity \citep[see also e.g.][]{McIntoshDePontieu2012}. Similarly, \citet{PantVaibhav2019} investigate the discrepancy between the observed wave energy, calculated from the Doppler velocities, and the true wave energy and find that the additional energy is present in the non-thermal line widths. Figure \ref{av_lw_s2} shows the evolution of the average line width, calculated using the full-width-half-maximum (FWHM) of the specific intensity, for simulation S2. Since the average temperature during the simulation does not change significantly over time, the thermal line width remains approximately constant throughout the simulation. Hence, the increase in the line width confirms that some of the `lost' energy is indeed hidden in the non-thermal line width.

\section{Discussion and conclusions} \label{sec_Discussion}

In this paper, we have examined the synthetic emission data for two 3D MHD simulations which model the propagation and interference of transverse waves in complex magnetic fields, where the two simulations differ in the complexity of the initial background field. Waves are excited by a sinusoidal, transverse driver imposed at the bottom boundary and over time, are distorted by complex interference and phase mixing.

For LOS1 and LOS2 (Fig. \ref{los_angles} and Table \ref{los_angle_table}), the total intensity largely corresponded to the temperatures in the domain. The Fe \rom{16} line captures the hotter central column of plasma, where the strongest braiding of the magnetic field is present, as well as locations of adiabatic heating. The cooler Fe \rom{12} line essentially detected the `inverse' of Fe \rom{16}, i.e.~the cooler outer regions of the simulation domain and, as the simulation progressed, locations of rarefaction. The magnetic field complexity was not associated with similar intricate structuring in the intensities along LOS1 and LOS2. The only evidence of complexity in the magnetic field was identified by examining the running differences along the magnetic field structure (LOS3 and 4). Regions of complex field could be detected using careful analysis of multiple emission lines but even then, an absolute comparison of the level of fine-scale structuring in the two simulations was not possible. In the S1 simulation (less braiding), tracing the magnetic field lines showed that one of the flux tubes is still distinguishable in the initial setup. However, with the exception of the Fe \rom{16} line along LOS3, the integrated intensities showed little evidence of the presence of this structure.

Despite the incompressible nature of the boundary driver, due to non-linearity and coupling to compressive wave modes, the waves are detectable in the LOS intensities. For S1, there are indeed (weak) signatures in the intensity and running difference (of the intensities) in both Fe lines. However, for the more complex field in S2, the wave front is substantially deformed and compressibility occurs on smaller spatial scales, leading to increased cancellation of the density and temperature perturbations along the LOS. As a result, the wave propagation is barely detectable and could easily be mistaken for background noise. In addition to the transverse waves, (compressible) slow waves also enter the domain. Here, we found additional structuring in the wave fronts in the horizontal direction compared to the transverse waves. These shorter horizontal spatial scales of the slow wave were most easily visible in the intensity base differences and are caused by the  interplay between the direction of the driver and the direction of the local magnetic field. Of course, a different driver and/or local magnetic field topology would lead to different horizontal structuring.

During the first transit of the wave through the domain, we see an apparent rotational motion in the intensity running difference along LOS3, even though none are actually present in the domain. However, changing the LOS angle by just \ang{10}, results in this apparent motion being much harder to detect, although the effect of a minor change to the viewing angle on the observations may be overcome by removing the average background propagation of the running difference structure. This apparent rotation also vanishes once the waves reflect off the top boundary and the wave interference results in a random chaotic pattern in the intensity running differences. It is therefore unlikely that this apparent rotational motion would be an issue in the interpretation of waves in closed magnetic structures, particularly given the presence of wave interference. However, in open field regions or where wave propagation is largely unidirectional (e.g. due to strong damping along the structure or efficient transmission at one of the footpoints), distinguishing transverse and rotational motions might not always be trivial and it is useful to be aware of the potential for misinterpretation when analysing wave propagation through complex magnetic field structures.

The strongest signature of the transverse waves was present in the Doppler velocities. For the more complex field, we see more phase mixing, evident from the distortion of the wave fronts \citep{HowsonDeMoortel2020}. This was especially visible in the hotter Fe \rom{16} line, since the hotter plasma corresponds to regions of more complex magnetic field. After the waves reflect off the top boundary, the combination of phase mixing and wave interference of upward and downward propagating waves throughout the domain leads to extensive distortion of the wave fronts. The Doppler shift patterns become increasingly fine-structured and chaotic, complicating the interpretation of the nature of the wave mode. Indeed, at certain instances along LOS1, the Doppler shifts could again be misinterpreted as torsional motions. Similar side-by-side red-blue signatures were also present along LOS2 once the polarisation of the wave changed to a mixture of $v_y$ (the direction of the boundary driven waves) and $v_x$. Comparable Doppler velocity profiles are identified in \citet{GoossensSoler2014} although for a different numerical set up. In particular, the authors investigate the synthetic emission from an over dense plasma cylinder in a uniform magnetic field, with an imposed velocity field replicating the behaviour of the kink wave. Observations along the direction of the transverse motion show red shift in the location of the internal plasma and blue shift in the external plasma. This red-blue structure is similar to our observations along LOS1. They also examine the LOS perpendicular to the transverse motion of the kink wave. Again, red-blue Doppler shift structures are formed. However, this time they appear to be the signatures of a directly driven torsional \Alfven wave, rather than simply the azimuthal component of a kink mode.



Using the synthetic spectroscopic data, we estimated the kinetic energy in the simulations. We considered the effects of various factors, including the LOS angle, magnetic field complexity and the emission lines (Fe \rom{12} and Fe \rom{16}). In the most optimal configuration, where the LOS is parallel to the boundary driver (LOS1) and the magnetic field is less braided (S1), we achieved reasonably accurate estimates for the total kinetic energy. However, for the least optimal configuration, namely a LOS perpendicular to the driver (LOS2 in the S1 case), we attain estimates several orders of magnitude less than the physical value. In practice, estimates of the total kinetic energy are likely to be between these two cases and could underestimate the true kinetic energy by an order of magnitude, where some of this energy will be represented by enhanced non-thermal line widths \citep[e.g.][]{McIntoshDePontieu2012, PantVaibhav2019}. Of course, additional uncertainties in the density profile and the LOS integration length may further increase the uncertainty in the energy measurements. In a previous study, \citet{DeMoortelPascoe2012} use a simple 3D numerical model of wave propagation along multiple loop strands to examine the effect of LOS integration on estimating the energy budget. They use a more complex driver than the one presented in this article, which is designed to mimic random footpoint motions. This may also have an effect on the energy budget, given that we found substantially different estimates from a LOS parallel to a LOS perpendicular to our unidirectional driver (i.e. a simple back and forth driver along one direction, in this case the $y$-axis). In their study, the authors find that the LOS kinetic energy budget captures, at best,  40\% of the actual kinetic energy generated, suggesting that the LOS integration has a substantial impact on the kinetic energy estimated from Doppler shifts.

In summary, many of the observables we have discussed in this paper are relatively easily understood, at least with the knowledge of the evolution of the physical parameters in the 3D domain. However, our analysis does highlight that care is required in some instances, where the propagation of a simple wave front in a complex magnetic field can lead to complex patterns in the observables. In particular, we found that when considering the intensities or Doppler velocities in isolation, distinguishing between transverse and rotational motions is not always trivial. Finally, we investigated the estimated kinetic energy which is highly dependent on the spatial scale of the wave driver, the polarisation of the wave, complexity of the background field and the LOS angles.

\vspace{1cm}
{\emph{Acknowledgements.}} The authors would like to thank Paola Testa, 
Jorrit Leenaerts and Vaibhav Pant for helpful discussion on the double-Gaussian fitting routines. We would also like to thank Patrick Antolin and Tom Van Doorsselaere for assistance with the FoMo code. The research leading to these results has received funding from the UK Science and Technology Facilities Council (consolidated grants ST/N000609/1 and ST/S000402/1) and the European Union Horizon 2020 research and innovation programme (grant agreement No. 647214). IDM acknowledges support from the Research Council of Norway through its Centres of Excellence scheme, project number 262622.

\bibliographystyle{aa}        
\bibliography{fomo_waves_paper.bib}           

\end{document}